\documentclass[]{emulateapj}
\usepackage{psfig}
\usepackage{natbib,graphicx}

\def\boldxi{\mbox{\boldmath $\xi$}}

\def\boldOmega{\mbox {\boldmath $\Omega$}}
\def\boldz{\mbox {\boldmath $z$}}
\def\boldr{\mbox {\boldmath $r$}}
\def\bnabla{\mbox{\boldmath $\nabla$}}
\def\boldr{\mbox{\boldmath$r$}}
\def\boldez{\mbox {\boldmath $e_z$}}
\def\boldeh{\mbox {\boldmath $e_h$}}
\def\boldk{\mbox{\boldmath$k$}}
\def\boldv{\mbox{\boldmath$v$}}
\def\div{\bnabla\cdot}
\def\divxi{\bnabla\cdot{\boldxi}}

\def\Bruntfreq{Brunt-V{\"a}is{\"a}l{\"a}\,\,\,}
\def\refnew#1{(\ref{#1})}
\def\be{\begin{equation}}
\def\ee{\end{equation}}
\def\ba{\begin{eqnarray}}
\def\ea{\end{eqnarray}}

\def\km{\, \rm km}

\let\pomega\varpi

\begin{document} 
	
\title{\mbox{Origin of Tidal Dissipation in Jupiter:
I. Properties of Inertial-Modes}}

\lefthead{Tides in Jupiter}
\righthead{Wu}

\author{Yanqin Wu}
\affil{Department of Astronomy and Astrophysics, 60 St. George Street, 
	University of Toronto, Toronto, ON M5S 3H8, Canada}

\begin{abstract}
We study global inertial-modes with the purpose of unraveling the role
they play in the tidal dissipation process of Jupiter. For spheres of
uniformly rotating, neutrally buoyant fluid, we show that the partial
differential equation governing inertial-modes can be separated into
two ordinary differential equations when the density is constant, or
when the density has a power-law dependence on radius. For more
general density dependencies, we show that one can obtain an
approximate solution to the inertial-modes that is accurate to the
second order in wave-vector.  Frequencies of inertial-modes are
limited to $\omega < 2 \Omega$ ($\Omega$ is the rotation rate),
with modes propagating closer to the rotation axis having higher
frequencies. An inertial-mode propagates throughout much of the sphere
with a relatively constant wavelength, and a wave amplitude that
scales with density as $1/\sqrt{\rho}$. It is reflected near the
surface at a depth that depends on latitude, with the depth being much
shallower near the special latitudes $\theta = \cos^{-1}
\pm \omega/2\Omega$.  Around this region, this mode has the highest
wave amplitude as well as the sharpest spatial gradient (the
``singularity belt''), thereby incurring the strongest turbulent
dissipation.  Inertial-modes naturally cause small Eulerian density
perturbations, so they are only weakly coupled to the tidal potential.
In a companion paper, we attempt to apply these results to the problem
of tidal dissipation in Jupiter.
\end{abstract}




\section{INTRODUCTION}
\label{sec:nl-intro}

\subsection{Physical Motivation}
\label{subsec:observation}

We study properties of global inertial-modes in neutrally buoyant,
rotating spheres. We briefly introduce our motivation for doing so,
but refer readers to \citet{gordon}, as well as
\citet[][hereafter {\bf Paper II}]{2004b} for a more thorough
introduction.

Jupiter's tidal dissipation factor (dimensionless $Q$ value) has been
estimated to be $10^5 \leq Q \leq 2\times 10^6$
\citep{goldreichsoter,peale}, 
based on the current resonant configuration of the Galilean
satellites. $Q$ is inversely proportional to the rate of tidal
dissipation.  So far, the physical explanation for this $Q$ value has
remained elusive, with the most reliable calculations yielding a tidal
$Q$ orders of magnitude greater than the above inferred
value. Importantly, known extra-solar planets also exhibits $Q$ value
of the same order as Jupiter \citep{wu03}. This is inferred from the
value of the semi-major axis ($\sim 0.07 AU$) within which extra-solar
planets are observed to possess largely circularized orbits.

We call attention to two common characteristics shared by Jupiter and
the close-in exo-planets: except for a thin radiative atmosphere, the
interior of these bodies are convective; and they spin fast -- typical
spin periods are shorter or comparable to twice the tidal forcing
period. This motivates us to search for an answer to the tidal $Q$
problem that relates to rotation.  In a sphere of neutrally buoyant
fluid, rotation gives rise to a branch of eigenmodes:
inertial-modes. These modes are restored by Coriolis force and have
many interesting properties that make them good candidates for
explaining the tidal $Q$ value.

\subsection{Previous works}
\label{subsec:past}

Waves restored by Coriolis force have been studied extensively in
the oceanographic and atmospheric sciences. Various names are associated
with them, e.g., Rossby waves, planetary waves, inertial waves,
R-modes \citep{greenspan}. In these contexts, the waves are
typically assumed to propagate inside a thin spherical shell.

There have been a few studies of
inertial-modes\footnote{Inertial-modes (also called generalized
r-modes, hybrid modes) refer to rotationally restored modes in
zero-buoyancy environment.} in the astrophysical
context. \citet{schenk} gives a fairly complete survey of the
literature related to inertial-modes. We refer readers to that paper
for a better understanding of the nomenclature and past efforts. Here,
we only mention a few related early works that have particular impact
on the current one.

\citet{bryan} studied tidal forcing of oscillations in a
rotating, uniform density spheroid (or ellipsoid). He performed a
coordinate transformation (pioneered by Poincar{\'{e}}) under which
the oscillation equation becomes separable. This immensely facilitates
ours and others' study of inertial-modes.
\citet{lindblom} followed essentially the same approach.


\citet{papaloizou2} studied inertial-modes in a
fully convective star with an adiabatic index $\Gamma = 5/3$
(equivalent to a $n=1.5$ polytrope). They realized that one could
obtain the eigenfrequency spectrum fairly accurately without detailed
knowledge of the eigenfunction. This is achieved using the variational
principle and by expanding the eigenfunction in a well-chosen basis.
They also pointed out that the fully convective case is special in
that gravity-modes vanish so inertial-modes form a single sequence in
frequency that depends only on rotational frequency.

\citet[][{\bf LF} from now on]{lockitch} studied
inertial-modes in spheres with arbitrary polytropic density
profile. They expressed the spatial structure of each inertial-mode as
a sum of spherical harmonic functions and curls of the spherical
harmonics. They presented some eigenfrequencies for, e.g., $n=1$
polytrope. We will compare our results against theirs.

Inertial-modes have also been attacked numerically, via integration of
the characteristics \citep{rieutord, gordon}, the finite difference
method \citep{savonije2}, and the spectral method \cite{gordon}.

\subsection{This Work}
\label{subsec:thiswork}

There are two purposes to this paper. First, it lays down the
foundation for Paper II where we discuss our attempt to solve the
tidal $Q$ problem. Second, it presents a new series of exact solutions
to inertial-modes in spheres with power-law density profile, as well
as approximate solutions for spheres with an arbitrary (but smoothly
varying) density profile.

Our approach centers on the ability to reduce the partial differential
equation governing fluid motion in a rotating sphere to ordinary
differential equations. This semi-analytical approach produces results
that are both easily reproducible and have clear physical
interpretations. The mathematics are concentrated in \S
\ref{sec:structure}.  We then expose properties of inertial-modes that
are relevant for its interaction with the tidal perturbations (\S
\ref{sec:property}). A discussion section (\S
\ref{sec:discussion}) follows in which we consider the validity 
of our assumptions.  

Readers interested in the tidal dissipation problem alone are referred
to \S \ref{sec:summary} for a brief summary of the features of
inertial-modes upon which we build our theory of tidal dissipation
(Paper II).

Readers interested in getting only a favor of how inertial-modes look
like are referred to Figs. \ref{fig:plota} \&
\ref{fig:density-surf}. The accompanying description (\S \ref{subsec:graphic}) 
as well as a comparison between inertial-modes and the more commonly
known gravity- and pressure-modes (\S \ref{subsec:differing}) may prove
useful as well.

\section{Inertial-Mode Eigenfunctions}
\label{sec:structure}

This section documents our effort in obtaining semi-analytical
eigenfunctions for inertial-modes. The relevant equation of motion is
first introduced in \S \ref{subsec:equation}. We then deal with
increasingly more complicated and more realistic cases. This proceeds
from simple uniform density spheres (\S
\ref{subsec:constantrho}), to power-law density spheres (\S
\ref{subsec:powerlaw}), and lastly to spheres with realistic planetary 
density profiles (\S \ref{subsec:approximate})

\subsection{Equations of Motion}
\label{subsec:equation}

Consider a planet spinning with a uniform angular velocity of
$\boldOmega$ pointing in the $\boldz$ direction. In the rotating
frame, the equations for momentum and mass conservation read
\begin{eqnarray}
\rho \ddot{\boldxi} + 2 \rho \boldOmega\times \dot{\boldxi} & = & 
- \bnabla p^\prime + {{\bnabla p}\over{\rho}}\, \rho^\prime - \rho
\bnabla \Phi^\prime,
\label{eq:eqnmotion}\\
\rho^\prime + \div(\rho\boldxi) & = & 0,
\label{eq:mass}
\end{eqnarray}
where $\boldxi$ is the displacement vector, while $p^\prime$,
$\rho^\prime$ and $\Phi^\prime$ are the Eulerian perturbations to
pressure, density and gravitational potential, respectively. We ignore
rotational deformation to the hydrostatic structure, as well as the
centrifugal force associated with the perturbation. Both these terms
are smaller by a factor $\sim (\Omega/\Omega_{\rm max})^2$ than the
terms we keep, with $\Omega_{\rm max}$ being the break-up spin-rate of
the planet. This factor is $\sim 10\%$ for Jupiter and much smaller for
close-in extra-solar planets which likely have reached
spin-synchronization with the orbit.

We restrict ourselves to adiabatic perturbations, hence the Lagrangian
pressure and density perturbations are related to each other by
\be
{{\delta p}\over{p}} = \Gamma_1 {{\delta \rho}\over{\rho}},
\label{eq:adiabatic}
\ee
where the adiabatic index $\Gamma_1 = \partial \ln p/\partial \ln
\rho|_s$ and is related to the speed of sound by $\Gamma_1 = c_s^2\,\rho/p$.

The interiors of giant planets are convectively unstable, with a low
degree of super-adiabaticity, as is guaranteed by the fact that the
convective velocity is fairly subsonic. This allows us to treat the
fluid as neutrally buoyant. Setting the \Bruntfreq frequency to zero,
we obtain,
\be
{{d\rho}\over{dr}} = {{\rho}\over{\Gamma_1 p }} {{dP}\over{dr}}.
\label{eq:zerobrunt}
\ee

Given the following expression which relates the Lagrangian and the
Eulerian perturbations in a quantity $X$,
\be
\delta X = X^\prime + \boldxi \cdot \bnabla X,
\label{eq:lageuler}
\ee
equations \refnew{eq:adiabatic} and \refnew{eq:zerobrunt} combine to
yield
\be
{{p^\prime}\over{p}} = \Gamma_1 {{\rho^\prime}\over{\rho}},
\label{eq:ptorho}
\ee
and the right-hand side of equation
\refnew{eq:eqnmotion} can be simplified into
\be
 - {{\bnabla p^\prime}\over{\rho}} + {{\bnabla p}\over{\rho}}\,
 {{\rho^\prime}\over{\rho}} - \bnabla \Phi^\prime = -
 \bnabla\left({p^\prime \over{\rho}}\right) - \bnabla \Phi^\prime = 
- \bnabla \omega^2 \psi,
\label{eq:definepsi}
\ee
where we have introduced a new scalar 
\be
\psi = {1\over \omega^2} \left( {{p^\prime}\over \rho} + \Phi^\prime \right) =
{1\over \omega^2} \left( c_s^2 {{\rho^\prime}\over \rho} + \Phi^\prime \right),
\label{eq:definepsi2}
\ee
with $\omega$ being the mode frequency in the rotating frame. We adopt
for all variables the following dependence on time ($t$) and on the
azimuthal angle ($\phi$): $X \propto
\exp{[i(m\phi-\omega t)]}$.  Modes with denotation $(m,\omega)$ and $(-m, -\omega)$ are
physically the same mode, so we restrict ourselves to $\omega \geq 0$,
with $m > 0$ representing a prograde mode, and $m< 0$ a retrograde
one.

In solving for the inertial-mode eigenfunction, we adopt the Cowling
approximation (negligible potential perturbations associated with the
fluid movement) and ignore any external potential forcing (e.g., the
tidal potential), so $\Phi^\prime = 0$. The former is justified in \S
\ref{subsec:assumption}, and the latter is relevant as we are interested in free 
oscillations. Equation \ref{eq:definepsi2} yields,
\be
\rho^\prime = {{\omega^2 \rho}\over{c_s^2}} \psi.
\label{eq:rhoprime}
\ee

Following convention, we define the following two dimensionless
numbers:
\be
\mu \equiv {{\omega}\over{2\Omega}}, \hskip1.0in q = {1\over \mu}.
\label{eq:definemu}
\ee
Inertial-modes satisfy $0 < \mu \leq 1$. Equations
\refnew{eq:eqnmotion}-\refnew{eq:mass} can now be recast as
\begin{eqnarray}
\boldxi + i q\, (\boldez \times \boldxi) & = & 
\bnabla \psi, 
\label{eq:xi1}\\
\divxi + {{\omega^2}\over{c_s^2}} \psi & = & {{{\mbox {\boldmath
$e_r$}} \cdot \boldxi}\over{H}} = {g\over{c_s^2}} \,
({\mbox {\boldmath $e_r$}} \cdot \boldxi),
\label{eq:xi2}
\end{eqnarray}
where $H \equiv - dr/d\ln \rho$ is the density scale height and $H =
c_s^2/g$ (eq. [\ref{eq:zerobrunt}]), with $g$ being the local
gravitational acceleration.

Operating on equation \refnew{eq:xi1} with $\boldez \cdot$ and
$\boldez\times$, respectively, and replacing $\boldez\cdot \boldxi$
and $\boldez\times \boldxi$ using the resultant equations, we obtain
the following relationship between $\boldxi$ and $\psi$
\begin{eqnarray}
\boldxi = 
{1\over{1-q^2}} (1 - i q\, \boldez {\bold \times}\, - q^2 \boldez \, \boldez
\cdot) \, \,\bnabla \psi.
\label{eq:xipsi}
\end{eqnarray}
We substitute this equation into equation \refnew{eq:xi2} to eliminate
$\boldxi$ and to acquire the following key equation for $\psi$,
\begin{eqnarray}
& & \nabla^2 \psi - q^2 {{\partial^2 \psi}\over{\partial z^2}} =
\nonumber \\ & & {1\over H}
\left( {{\partial \psi}\over{\partial r}} - q^2 \cos\theta {{\partial 
\psi}\over{\partial z}} - {{mq}\over r} \psi\right) -  (1-q^2) 
{{\omega^2}\over{c_s^2}} \psi.
\label{eq:psifull}
\end{eqnarray}
Here $\theta$ is the zenith angle and $\cos\theta = z/r$ with $z$
being the height along the rotational axis. Let $\pomega$ be the
cylindrical radius.  The partial derivatives here are to be understood as
$\partial/\partial r =
\left.\partial/\partial r\right|_\theta$, $\partial/\partial \theta =
\left.\partial/\partial
\theta\right|_r$, $\partial/\partial z= \left.\partial/\partial
z\right|_{\pomega}$, and $\partial/\partial\pomega =
\left.\partial/\partial\pomega\right|_z$.  In general, the above
partial differential equation is not separable in any coordinates and
only fully numerical solutions could be sought.

The displacement vector in spherical coordinates is related to $\psi$
as
\refnew{eq:xipsi},
\begin{eqnarray}
\xi_r & = & {1\over{1-q^2}} \left( {{\partial\psi}\over{\partial r}} - 
{{mq}\over r} \psi - q^2 \cos\theta {{\partial \psi}\over{\partial z}}\right),
\label{eq:xir}\\
\xi_\theta & = & {1\over{1-q^2}} \left( {1\over r} 
{{\partial\psi}\over{\partial\theta}} - {{mqz}\over{r \pomega}} \psi + q^2
\sin\theta {{\partial\psi}\over{\partial z}}\right),
\label{eq:xitheta}\\
\xi_\phi & = & {1\over{1-q^2}} \left( {{i m}\over{\pomega}}\psi - i q {{\partial
\psi}\over{\partial \pomega}}\right).
\label{eq:xiphi}
\end{eqnarray}
In the following discussions (except where noted), we scale all
lengths by the radius of the planet ($R$). 

Equation \refnew{eq:psifull} describes the pulsation for both
inertial-modes and pressure-modes in a uniformly rotating,
neutrally-buoyant fluid -- gravity-modes have zero frequencies in such
a medium. While the compressional term ($(1-q^2)\omega^2/c_s^2 \psi$)
represents the main restoring force for the pressure-modes, it is
negligible (in comparison to the Coriolis terms) for inertial-modes as
the latter are much lower in frequency. We adhere to this
simplification throughout our analysis, and further justify it in
\S \ref{subsec:assumption}.

We now introduce an important coordinate transform which facilitates
much of our analysis. We follow \citet{bryan} in adopting a set of
ellipsoidal coordinates $(x_1, x_2$) which depend on the value of
$\mu$.  More details on these coordinates are presented in
Appendix
\ref{sec:coordinate}. Fig. \ref{fig:x1x2} depicts the topographic 
curves of $(x_1, x_2)$ in a meridional plane, as well as the curves of
constant $r$ in the $x_1 - x_2$ plane, when $\mu = 0.75$. Generally,
the $x_1$ coordinate is parallel to the $\pomega$ axis, while the
$x_2$ coordinate is parallel to the $z$ axis over much of the sphere.
On the spherical surface, either $x_1$ or $x_2$ (or both at the
special latitude $\theta = \cos^{-1} \pm \mu$) equals $\mu$; $x_1 = 1$
at the rotation axis and $x_2 = 0$ at the equator.

The ellipsoidal coordinates are a natural choice for studying
inertial-modes in a rotating sphere 
Motion restored by the Coriolis force (inertial-waves) would have liked
to follow the cylindrical coordinates, were it not for the spherical
topology of the object within which they dwell. The ellipsoidal
coordinates, being a hybrid between the cylindrical and the spherical
coordinates, are therefore particularly suitable for this purpose.

With this set of coordinates, the left-hand side of equation
\refnew{eq:psifull} is transformed into,
\begin{eqnarray}
& & \left(\nabla^2 - q^2 {{\partial^2 }\over{\partial z^2}}\right)\psi
\nonumber \\
&= & \left[ {1\over{\pomega}} {{\partial}\over{\partial\pomega}} \left(\pomega 
{{\partial}\over{\partial\pomega}}\right) + {1\over{\pomega^2}}
{{\partial^2}\over{\partial \phi^2}} + (1-q^2)
{{\partial^2}\over{\partial z^2}}\right]\, \psi \nonumber\\
&=  & 
{{1-\mu^2}\over{x_1^2 - x_2^2}} \left\{ 
\left[ (1-x_1^2) {{\partial^2}\over{\partial x_1^2}} - 
2 x_1 {{\partial}\over{\partial x_1}} - {{m^2}\over{1-x_1^2}}\right] - \right. 
\nonumber \\
& &  \left. \left[ (1-x_2^2) {{\partial^2}\over{\partial x_2^2}} - 
2 x_2 {{\partial}\over{\partial x_2}} -
{{m^2}\over{1-x_2^2}}\right]\right\}\, \psi .
\label{eq:seperatepsi}
\end{eqnarray}
This, as we shall see shortly, allows the separation of variables under
certain circumstances.

\subsection{Uniform Density Sphere}
\label{subsec:constantrho}

\subsubsection{Formal Solution}
\label{subsubsec:formal}

In a uniform density sphere, the scale height $H = \infty$ and the
right-hand side of equation \refnew{eq:psifull} vanishes. Its
left-hand side (eq [\ref{eq:seperatepsi}]) can benefit from the
following decomposition,
\be
\psi =  \psi_1 (x_1) \, \psi_2 (x_2),
\label{eq:psi12}
\ee
with $\psi_i$ satisfying
\be
{\cal D}_i \psi_i + K^2 \psi_i = 0.
\label{eq:psii}
\ee
Here the differential operator ${\cal D}_i$ is
\be
{\cal D}_i = {{\partial}\over{\partial x_i}} \left[(1-x_i^2) 
{{\partial}\over{\partial x_i}}\right] - {{m^2}\over{1-x_i^2}},
\label{eq:defineDi}
\ee
and $K$ is a constant introduced when we separate variables. 

This result was first obtained by \citet{bryan} and its solutions are
called 'Bryan's modes'.
In fact, the solutions to $\psi_1$ and $\psi_2$ are the associated
Legendre polynomials. Requiring $\psi$ to be finite at the rotation
axis ($x_1 = 1$), we find $\psi_1$ and $\psi_2$ to be the same spherical
harmonic of the first kind
\citep{abramowitz}: $\psi_1 =
\psi_2 = P_\ell^m(x)$ with $\ell$ being an integer, $K^2 =
\ell(\ell+1)$, and the variable $x$ taken over the ranges $x_1 \in
[\mu,1]$, and $x_2 \in [-\mu,\mu]$, respectively. We explicitly
require that $\psi_1 (x_1 =
\mu) = \psi_2 (x_2 = \mu)$ so the eigenfunction needs only one
normalization constant.

The following boundary conditions apply.  First, at the
equator ($x_2 = 0$), even-parity modes
\footnote{It is straightforward to show that the displacement vector
$\boldxi$ has the same equatorial symmetry as $\psi$.} satisfy
\be
\left.{{d\psi_2}\over{dx_2}}\right|_{x_2 = 0} = 0,
\label{eq:boundary2}
\ee
while odd-parity modes satisfy
\be
\psi_2|_{x_2=0} = 0.
\label{eq:boundary3}
\ee
Properties of the Legendre polynomials 
\citep{abramowitz} require that $(\ell+m)$ to be an even integer
in the former case, and odd in the latter.  Second, $\psi_1$ is finite
at the polar axis ($x_1 =1$). The numerical equivalent to this
statement is best realized by introducing a 
variable $g_i$ which is related to $\psi_i$ as
\be
\psi_i (x_i) = (1-x_i^2)^{|m|/2} g_i (x_i).
\label{eq:psig}
\ee
This variable satisfies (eq. [\ref{eq:psii})
\be
(1- x_i^2) {{d^2 g_i}\over{dx_i^2}} - 2x_i (|m|+1) {{dg_i}\over{dx_i}}
+ \lambda^2 g_i = 0,
\label{eq:eqng}
\ee
where $\lambda^2 = K^2 - |m|(|m|+1) = \ell(\ell+1)-|m|(|m|+1)$.
Regularity of the eigenfunction at $x_1 = 1$ then translates into
a boundary condition
\be
\left. {{d g_1}\over{d x_1}}\right|_{x_1 = 1} 
= {{\lambda^2}\over{2(|m|+1)}} g_1. 
\label{eq:boundary4}
\ee
Our solutions show that near the rotation axis, $g_1$ approaches a
constant,
while $\psi_1$ approaches zero (if $|m| > 0$).

In this problem, the eigenfunction can be solved independently of the
eigenvalue $\mu$. So to determine $\mu$, we need one more boundary
condition. We enforce the physical condition that there is vacuum
outside the planetary surface ($r=1$), and therefore pressure
perturbation at the surface has to be zero. Written in a convenient
form of $\delta p/\rho = 0$ \citep{unno}, this corresponds to (using
eq. [\ref{eq:xi2}] and ignoring the compressional term),
\be
{{\delta p}\over \rho} = \Gamma_1 {{\delta\rho}\over{\rho}}
{p\over\rho} = - \Gamma_1 (\divxi) {p\over\rho} = - g \xi_r = 0.
\label{eq:dprho}
\ee
We relate $\xi_r$ to $\psi$ using equation \refnew{eq:xir}, as well as
relations presented in Appendix \ref{sec:coordinate},
\begin{eqnarray}
\xi_r  =  & & {1\over{(1-\mu^2) (x_1^2 - x_2^2)(1-x_1^2)(1-x_2^2)}} \nonumber \\
& & \times \left[ x_1 {{\partial \psi}\over{\partial x_1}} {{\mu^2 -
x_2^2}\over{1-x_2^2}} - x_2 {{\partial \psi}\over{\partial x_2}}
{{\mu^2-x_1^2}\over{1-x_1^2}} \right.
\nonumber \\
& & \left. + m\mu
{{x_1^2-x_2^2}\over{(1-x_1^2)(1-x_2^2)}} \psi \right].
\label{eq:xirexp}
\end{eqnarray}
So requiring $\xi_r = 0$ at the surface ($x_1 = \mu$ or $|x_2| = \mu$) is
equivalent to requiring that
\ba
\left.{{d\psi_1}\over{dx_1}}\right|_{x_1=\mu}
& = & - {m\over{1-\mu^2}} \left. \psi_1\right|_{x_1=\mu}, \nonumber \\
\left.{{d\psi_2}\over{dx_2}}\right|_{|x_2|=\mu}
 & = &  - 
{\mbox{SIGN}}[x_2]
\,  {m\over{1-\mu^2}} \left. \psi_2\right|_{|x_2|=\mu}.
\label{eq:boundary1}
\ea
Since $\psi_1$ and $\psi_2$ have the same functional form
($P_\ell^m(x)$), these two equations are in fact the same thing.
%
%
In actual numerical procedure, we solve for $g_i$ as opposed to
$\psi_i$. Equation \refnew{eq:boundary1} is modified for $g_i$ as,
\be
\left.{{d g_1}\over{d x_1}}\right|_{x_1 = \mu} = 
{{-(m-|m|\mu)}\over{1-\mu^2}} g_1|_{x_1=\mu}.
\label{eq:boundary5}
\ee
Using equation \refnew{eq:psig}, we find that $g_1$ is a polynomial of
order $(\ell - |m|)$.\footnote{One can express $g_i$ as $d^{|m|}
P_\ell^0(x)/d x^{|m|}$, and $P_\ell^0(x)$ is a polynomial of order
$\ell$.} For each $(\ell, m)$ pair, there can therefore be $(\ell -
|m|)$ roots satisfying this boundary condition, with half of them
having $\mu > 0$. Another way of phrasing this is that, for each
eigenfunction ($\psi(x_1,x_2) = P_\ell^{|m|}(x_1)\,
P_\ell^{|m|}(x_2)$), there are $(\ell - |m|)$ eigenfrequencies.
Roughly half of these are prograde modes ($m > 0$) and the other half
retrograde modes ($m < 0$).


Table \ref{tab:constantrho} presents some example eigenfrequencies for
inertial-modes in a uniform density sphere. Our results agree with
those presented in Tables 3-4 of LF and Table 1 of \citet{lindblom}.

\subsubsection{The Dispersion Relation}
\label{subsubsec:dispersion}


In the previous section, we have introduced various eigenvalues ($K$,
$\lambda$, $\ell$ and $m$), as well as the eigenfrequency $\mu$. What
is the geometrical meaning of these eigenvalues and what is the
dispersion relation for inertial-modes?

Here, we present a derivation that answers these questions. We first
convert the independent variable from $x_i$ to $\Theta_i =
\cos^{-1} x_i$ (to be differentiated with
the spherical angle ${\theta}$) so that equation \refnew{eq:eqng}
becomes,
\be
{{d^2 g_i}\over{d\Theta_i^2}} + (2 |m|+1)
{{\cos\Theta_i}\over{\sin\Theta_i}} {{dg_i}\over{d\Theta_i}} +
\lambda^2 g_i = 0.
\label{eq:eqng2}
\ee
Adopting a WKB approach, we express
$g_i \propto \exp(i\int k_\Theta d\Theta_i)$ with the wave vector
$k_\Theta = k_R+ i k_I$ where the real part $k_R
\sim {\cal O}(\lambda) \gg 1$ and the imaginary part $k_I \sim {\cal O}(1)$.
Substituting this into equation \refnew{eq:eqng2}, and equating terms
of comparable magnitudes, we find
\begin{eqnarray}
k_R & \approx & \lambda, \nonumber \\
k_I & \approx & \left(|m|+{1\over 2}\right)
{{\cos\Theta_i}\over{\sin\Theta_i}} = \left(|m|+{1\over 2}\right)
\,{{d\ln \sin \Theta_i}\over{d\Theta_i}}.
\label{eq:krki1}
\end{eqnarray}
Together with equation \refnew{eq:psig}, this gives rise to the
following approximate solution for $\psi_i$ in the WKB regime,
\be
\psi_i = P_\ell^m (x_i) 
\propto {1\over{|\sin\Theta_i|^{1/2}}} \cos(\lambda\Theta_i + \alpha).
\label{eq:psienv1}
\ee
So $\psi_i$ is an oscillating function with a roughly constant envelope.
Moreover, $\alpha = -\lambda\pi/2 + \pi$ for even-parity modes
(eq. [\ref{eq:boundary2}]); while $\alpha= -\lambda\pi/2 + \pi/2$ for
odd-parity modes (eq. [\ref{eq:boundary3}]). Nodes of $\psi_i = 0$ are
roughly evenly spaced in $\Theta \in [0,\pi/2]$ space, and the value
of $\lambda$ corresponds with twice the total number of nodes (more
below).

To obtain the dispersion relation, we insert the above approximate
solution into the boundary condition at the surface
(eq. [\ref{eq:boundary1}]) where $\Theta_0 = \cos^{-1} \pm \mu$. This
yields,
\be
\lambda \sin(\lambda\Theta_0+\alpha) 
+ {m\over{\sqrt{1-\mu^2}}} \cos(\lambda\Theta_0+\alpha) \approx 0.
\label{eq:bc0sin}
\ee
For $\lambda \gg 1$, this can be approximated by
\be
\sin(\lambda\Theta_0 + \alpha) \approx \lambda\Theta_0+\alpha+ n\pi
\approx {\pm m\over{\sqrt{1-\mu^2}\lambda}},
\label{eq:bc0sin2}
\ee
where $n$ is an integer. We retain only the positive sign as $m$ can
be positive or negative. For even-parity modes, the dispersion
relation for the eigenfrequency runs as
\begin{eqnarray}
\mu & = & 
\cos \Theta_0 
= \cos\left( - {{n\pi}\over{\lambda}} +
{m\over{\sqrt{1-\mu^2}\lambda^2}} + {\pi\over 2} -
{\pi\over{\lambda}}\right) \nonumber \\
& =  &
\sin\left( {{{(1+n)}\pi}\over{\lambda}} - {m\over{\sqrt{1-\mu^2}\lambda^2}} 
\right) ;
\label{eq:mueven}
\end{eqnarray}
while for odd-parity, it is
\begin{eqnarray}
\mu & = & 
\cos\left( - {{n\pi}\over{\lambda}} +
{m\over{\sqrt{1-\mu^2}\lambda^2}} + {\pi\over 2} -
{\pi\over{2\lambda}}\right) \nonumber \\ & = & \sin \left( {{{({1\over
2}+ n)}\pi}\over{\lambda}} - {m\over{\sqrt{1-\mu^2}\lambda^2}}
\right) .
\label{eq:muodd}
\end{eqnarray}

If we define the number of nodes in the $\Theta_1 \in [0,\Theta_0]$
range to be $n_1$, and the number in the $\Theta_2 \in
[\Theta_0,\pi/2]$ range to be $n_2$, they satisfy $n_1
\pi \approx \lambda
\Theta_0$, $n_2 \pi \approx \lambda (\pi/2 -
\Theta_0)$. So $\lambda \approx 2 (n_1 + n_2)$, as we have mentioned earlier.
Moreover, equation \refnew{eq:bc0sin2} yields $n
\approx n_2 - 1 + m/\sqrt{1-\mu^2}\lambda\pi$ 
for even modes, and $n \approx n_2 - 1/2 + m/\sqrt{1-\mu^2}\lambda\pi$
for odd modes. Substituting these estimates into equations
\refnew{eq:mueven} \& \refnew{eq:muodd}, we find the following
simplified dispersion relation,
\be
\mu = 
{\omega\over{2\Omega}} \approx
 \sin \left( {{n_2 \pi} \over{ 2 (n_1 + n_2)}} \right) \approx
 \cos \left( {{n_1 \pi} \over{ 2 (n_1 + n_2)}} \right) .
\label{eq:muevenodd}
\ee
This dispersion relation has also been numerically confirmed.  So for a
given set of $(\lambda,m)$, the eigenfrequency $\mu$ rises as the
partition of the $x_1/x_2$ space moves from $\mu = 0$ (when $n_1 \sim
\lambda/2$, $n_2 = 0$) to $\mu = 1$ (when $n_1 = 0$, $n_2 =
\lambda/2$), and there are  $\sim \lambda/2$ discreet eigenfrequencies.
Recall that $\lambda = \sqrt{\ell(\ell+1)-|m|(|m|+1)} \approx \ell -
|m| $. So for each $(\lambda,|m|)$ pair, we find $\sim (\ell-|m|)/2$
prograde eigenmodes and $\sim (\ell-|m|)/2$ retrograde ones,
consistent with the results found in
\S \ref{subsubsec:formal}.


In \S \ref{subsec:wkb}, we will discuss further the geometrical
meaning of the eigenvalues in spherical coordinates.

%

\subsection{Power-Law Density Sphere}
\label{subsec:powerlaw}

\subsubsection{The equation is Separable!}
\label{subsubsec:eqnsep}

We are inspired by the planetary density profile to investigate
inertial-modes in a sphere where density depends on radius as a
power-law. To our pleasant surprise, the equation turns out be
separable in this case. This forms the basis for much of our work when
dealing with real planets.

We adopt the following power-law density profile (see
eq. [\ref{eq:pomegar}])
%
\be
\rho 
\propto (1-r^2)^\beta =  \rho_0 
\left[ (x_1^2 - \mu^2) (\mu^2 - x_2^2) \right]^\beta.
\label{eq:powerrho}
\ee
This is not a usual choice and is different from the more typically
discussed polytropes. However, near the surface, our power-law density
profile behaves like a normal polytrope. Let $\delta=1-r$ be the depth
into the planet. For $\delta
\ll 1$, we find
\ba
\delta & = & 1 - r = 1 - \sqrt{1 - 
{{(x_1^2 - \mu^2)(\mu^2-x_2^2)}\over{(1-\mu^2)\mu^2}}} 
\nonumber \\
&\approx & {1\over 2}{{(x_1^2 -
\mu^2)(\mu^2-x_2^2)}\over{(1-\mu^2)\mu^2}}.
\label{eq:deltaexp}
\ea
So near the surface, equation \refnew{eq:powerrho} yields $\rho
\propto \delta^\beta$, as is the case for a polytrope model with 
the polytropic index $n = \beta$.  The advantage of choosing such a
profile will become clear soon.

We obtain the following expression for the density scale-height $H$:
\begin{eqnarray}
H^{-1} & \equiv & - {{d\ln \rho}\over{dr}} = - \left( {{z\over
r}{\partial\over{\partial z}} + {{\pomega\over r}{\partial
\over{\partial \pomega}}}} \right) \ln \rho \nonumber \\
& = & 2{{(1-\mu^2) \mu^2 \beta r}\over{(x_1^2 - \mu^2)(\mu^2 - x_2^2)}},
\label{eq:Hbeta}
\end{eqnarray}
where equations \refnew{eq:pomegar} \& \refnew{eq:partialr} are used.
Together with equations
\refnew{eq:partialpomega}-\refnew{eq:partialr}, this allows equation
\refnew{eq:psifull} to be recast into (again ignoring the compressional term),
\ba
& & \left[{\cal D}_1 + {{2\beta x_1 (1-x_1^2)}\over{x_1^2 - \mu^2}}
{\partial\over{\partial x_1}} + {{2 \mu \beta m}\over{x_1^2 -
\mu^2}}\right] \psi - \nonumber \\
& & \left[{\cal D}_2 + {{2\beta x_2 (1-x_2^2)}\over{x_2^2 - \mu^2}} 
{\partial\over{\partial x_2}} + {{2 \mu \beta m}\over{x_2^2 -
\mu^2}}\right] \psi  = 0,
\label{eq:psifull2}
\ea
where the operator ${\cal D}_i$ is defined in equation
\refnew{eq:defineDi}. So when density follows the power-law profile 
(eq. [\ref{eq:powerrho}]), the equation for the inertial-modes is
again {\bf separable}: $\psi(x_1,x_2) = \psi_1 (x_1) \psi_2 (x_2)$
with $\psi_i$ satisfying
\ba
& &\left[{\cal D}_i +  {{2\beta x_i (1-x_i^2)}\over{x_i^2 - \mu^2}} 
{d\over{d x_i}} + {{2 \mu \beta m}\over{x_i^2 -
\mu^2}}\right] \psi_i  + K^2 \psi_i  \nonumber \\
& & = {\cal E}_i\psi_i + K^2 \psi_i = 0,
\label{eq:psi2}
\ea
where we introduce a new operator ${\cal E}_i$.  Note that this
equation is exact (except for omitting the compressional term) for the
power-law density sphere.

Numerically, it is more accurate to solve for $g_i =
\psi_i/(1-x_i^2)^{|m|/2}$. This function satisfies
\ba
& & (1- x_i^2) {{d^2 g_i}\over{dx_i^2}} - 2x_i (|m|+1)
{{dg_i}\over{dx_i}} + {{2\beta x_i (1-x_i^2)}\over{x_i^2 - \mu^2}}
{{dg_i}\over{dx_i}} \nonumber \\ & & + \left[\lambda^2 - {{2\beta |m|
x_i^2}\over{x_i^2 - \mu^2}} + {{2 \beta m \mu }\over{x_i^2
-\mu^2}}\right] g_i = 0,
\label{eq:psig2}
\ea
where $\lambda^2 = K^2 - |m|(|m|+1)$. The boundary conditions
presented in \S
\ref{subsubsec:formal} apply here as well, with the exception of 
equation \refnew{eq:boundary4} which should be modified into
\be
\left.{{d g_1}\over{d x_1}}\right|_{x_1 = 1} = 
{{\lambda^2 +2\beta {{m\mu-|m|}\over{1-\mu^2}}}\over{2(|m|+1)}} g_1,
\label{eq:boundary6}
\ee
So unlike the uniform density case, both the inertial-mode equation
and the boundary conditions now depend explicitly on the value of
$\mu$ as well as the sign of $m$. This breaks the eigenfunction
degeneracy existing in the uniform density case: each eigenmode now
has a unique eigenfunction.

In Fig. \ref{fig:eigenfunction}, we plot the resulting eigenfunction
for a $m = - 2$ mode in a $\beta=1$ model. We also contrast the
density profile of our $\beta=1$ model with the conventional $n=1$
polytrope model ($p\propto \rho^2$). Details of how we construct the
$\beta$ models are presented in Appendix \ref{sec:betamodel}. 

Table \ref{tab:betarho} lists some sample eigenfrequencies for $|m|=2$
modes in various $\beta$ models. The number of eigenmodes remains
conserved when $\beta$ varies, with a close one-to-one correspondence
between modes in different density profiles. This makes mode typing
trivial. We also compare our results against those of LF obtained for
a $n=1$ polytrope.

\subsubsection{WKB Envelope}
\label{subsubsec:envelope}

\begin{figure*}
      \centering{
      \vbox{\psfig{figure=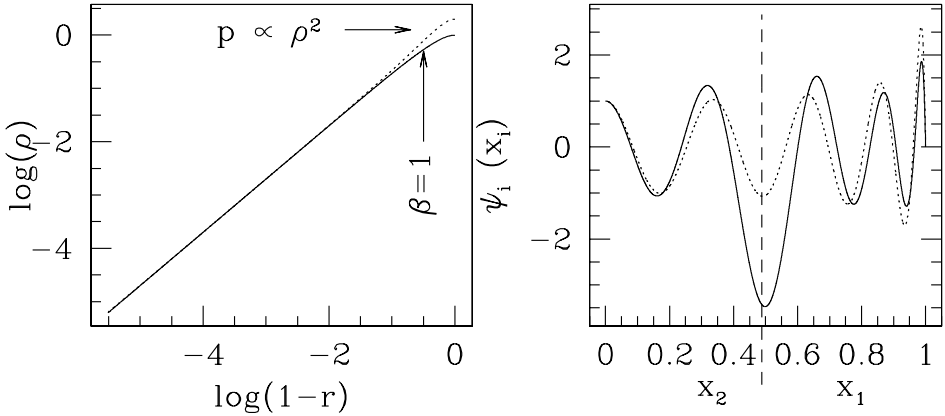,width=0.80\hsize,%
       bbllx=0pt,bblly=160pt,bburx=650pt,bbury=440pt,clip=}}
       }
\caption{Left-hand plot shows the logarithm of the density ($\log \rho$) 
as a function of the logarithm of the depth ($\log(1-r)$) for a
$\beta=1$ power-law model (solid curve). The density scale is
arbitrary. The dotted line is the density profile for a $n=1$
polytrope ($p\propto \rho^2$).  The two profiles behave similarly
except near the center.  Right-hand panel: eigenfunctions
$\psi_1(x_1)$ and $\psi_2(x_2)$ as a function of $x_1$ or $x_2$ for an
even-parity, retrograde $m=-2$, $\mu = 0.4881$ mode in the $\beta = 1$
model (the solid curve).  The global eigenfunction $\psi(x_1, x_2) =
\psi_1\, \psi_2$.  The dashed line demarcates the $x_1$ and $x_2$
boundary at $\mu = 0.4881$.  There are $5$ and $3$ nodes within $x_1$
and $x_2$ ranges, respectively, with a corresponding $\ell = 2(n_1 +
n_2)+|m| = 18$ (see Table
\ref{tab:betarho}).  As $x_i$ approaches $\mu$, the solid curve 
exhibits a rise in amplitude that is described by a $1/\sqrt{\rho}$
WKB-envelope. For comparison, we draw (in dotted curve) the
eigenfunction of a counterpart mode in the $\beta=0$ (uniform density)
model.}
\label{fig:eigenfunction}
\end{figure*}

While the eigenfunction $\psi$ inside a uniform density sphere has a
roughly constant envelope (eq. [\ref{eq:psienv1}]), $\psi$ in the
power-law density case behaves differently. Following the same
operation as in \S \ref{subsubsec:dispersion}, one finds
\begin{eqnarray}
k_R &\approx &\sqrt{\lambda^2 +
{{2\beta(m\mu-|m|\cos^2\Theta_i)}\over{|\cos^2\Theta_i -
\mu^2|}}}\approx
\lambda, \nonumber\\ 
k_I & \approx & 
 \left(|m|+{1\over2}\right){{\cos\Theta_i}\over{\sin\Theta_i}} -
{{\beta \cos\Theta_i \sin\Theta_i }\over{|\cos\Theta_i^2 - \mu^2|}}
\nonumber \\
&  = &  {d\over{d\Theta_i}}\left[ \ln \sin^{|m|+{1\over
2}} \Theta_i + \ln |\cos^2\Theta_i - \mu^2|^{\beta\over 2}\right],
\label{eq:krki2}
\end{eqnarray}
and
\be
\psi_i \approx {1\over{|\sin\Theta_i|^{1\over 2}\, |\cos^2\Theta_i - 
\mu^2|^{\beta\over 2}}} \cos(\lambda\Theta_i + \alpha).
\label{eq:psienv2}
\ee
So, the eigenfunction $\psi$ has an amplitude that rises toward the
surface (when $|x_i| \rightarrow \mu$). In fact, its envelope scales
with density as
\be
\psi = \psi_1 (x_1) \psi_2 (x_2) \propto {1\over{[(x_1^2 - \mu^2)
\, (\mu^2 - x_2^2)]^{\beta/2}}} \propto {1\over {\sqrt{\rho}}}.
\label{eq:psienv2b}
\ee
Fig. \ref{fig:eigenfunction} shows one example of such a WKB envelope.

We find that the dispersion relation is also modified slightly from
the case of a uniform density sphere. But qualitative features of
equation \refnew{eq:muevenodd} remain.

\subsection{Approximate Solution for Realistic Density Profiles}
\label{subsec:approximate}

The density profile inside a planet typically traces out two different
polytropes: near the surface, the gas can be approximated as an ideal
gas composed of diatomic molecules, with a mean degree of freedom 
$5$, so $\Gamma_1 ={\partial \ln P/\partial
\ln \rho}|_s = 7/5$,  and the correct polytrope number is $n \sim 2$;
in the interior of the planet, Coulomb pressure and electron
degeneracy modify the equation of state and raise $\Gamma_1$ to $\sim
2$, while reducing the polytrope number to $n \sim 1$. This motivates
us to model Jupiter's density profile using two power-laws of the form
in equation \refnew{eq:powerrho}: $\beta = 1$ in the interior and
$\beta = 1.8$ near the surface, with the transition occurring at a
radius $r
\approx 0.98$ (see Paper II). This reproduces profiles in realistic Jupiter models
\citep{guillotreview}.
The presence of a core or a phase transition complicates this
picture. We discuss them in more detail in Paper II.

In general, equation \refnew{eq:psifull} is not separable under these
density profiles. However, we discover an approximate solution which
is accurate to the second order in wave-numbers (${\cal O}
(1/\lambda^2)$ with $\lambda \gg 1$), for the case when the density
profile is power-law near the surface and smoothly varying in the
interior. This is largely inspired by the WKB result in \S
\ref{subsubsec:envelope}.

We first introduce a fiducial density $\rho_{\rm surf}$ which
satisfies equation \refnew{eq:powerrho} with a power-law index taken
to be that for the true density $\rho$ near the surface. We also
define $X = \rho_{\rm surf}/\rho$ so $X \approx 1$ near the surface
and deviates from unity toward the center. Inverse of the density
scale height can be written as
\be
{1\over H} = - {{d\ln \rho}\over{dr}} = {{d\ln X}\over{dr}} +
{{2(1-\mu^2) \mu^2 \beta r}\over{(x_1^2 - \mu^2)(\mu^2 - x_2^2)}}.
\label{eq:Hbeta2}
\ee
Introducing a variable $t$,
\be
t = (x_1^2 - \mu^2)(\mu^2-x_2^2) = \mu^2 (1-\mu^2)(1-r^2), 
\label{eq:definet}
\ee
one finds,
\be
{{d\ln X}\over{dr}} = - 2 r \mu^2 (1-\mu^2) {{d\ln X}\over{dt}}.
\label{eq:dlnxdt}
\ee

Now repeat the same calculations that lead to equation
\refnew{eq:psifull2}, we obtain
\ba
& & ({\cal E}_1 - {\cal E}_2)\psi - 2 {{d\ln X}\over{d\ln t}} 
\left[
{{(1-x_1^2) x_1}\over{(x_1^2 - \mu^2)}} {\partial\over{\partial x_1}}\right.
\nonumber \\
& & 
\left. + {{(1-x_2^2) x_2}\over{(\mu^2 - x_2^2)}} {\partial\over{\partial x_2}}
+ {{m \mu (x_1^2 - x_2^2)}\over t}
\right]\psi = 0,
\label{eq:psifull3}
\ea
where the operator ${\cal E}_i$ is defined in equation
\refnew{eq:psi2}. Here, since $\partial \psi /\partial x_i \sim \lambda \psi$, 
we can not ignore terms in the square parenthesis if we want to be
accurate to ${\cal O} (1/\lambda^2)$, the final goal of our procedure.

Instead, we experiment with the following decomposition for $\psi$,
\be
\psi = \sqrt{X} \psi_0(x_1, x_2) = \sqrt{{\rho_{\rm surf}}\over{\rho}} 
\psi_0 (x_1, x_2).
\label{eq:psiform3}
\ee
This is inspired by the WKB envelope presented in equation
\refnew{eq:psienv2b}. 

Formally expressing ${\cal E}_i \psi = a_i {\partial^2
\psi}/{\partial x_i^2} + b_i {\partial \psi}/{\partial x_i} + c_i
\psi$, where the meanings of $a_i, b_i$ and $c_i$ are clear from equation 
\refnew{eq:psi2}, we find
\be
{\cal E}_i \psi = \sqrt{X} {\cal E}_i \psi_0 + 2 a_i {{\partial
\sqrt{X}}\over{\partial x_i}} {{\partial \psi_0}\over{\partial x_1}} + 
\left( a_i {{\partial^2 \sqrt{X}}\over{\partial x_i^2}} + 
b_i {{\partial \sqrt{X}}\over{\partial x_i}}\right) \psi_0,
\label{eq:Epsi}
\ee
where
\be
{{\partial \sqrt{X}}\over{\partial x_i}} = {{x_i}\over{x_i^2 - \mu^2}}
\sqrt{X} {{d\ln X}\over{d\ln t}}.
\label{eq:Xx}
\ee
A straight-forward but lengthy derivation shows that equation
\refnew{eq:psifull3} can be recast into an equation for $\psi_0$,
\be
({\cal E}_1 - {\cal E}_2)\psi_0 + (d_1 - d_2) \psi_0 = 0,
\label{eq:psifull4}
\ee
where the coefficient $d_i$ is
\ba
d_i = & & {1\over{\sqrt{X}}}\left\{ {{\partial}\over{\partial
x_i}}\left[(1-x_i^2) {{\partial \sqrt{X}}\over{\partial x_i}}\right] +
{\left(2 \beta - 2 {{d\ln X}\over{d\ln t}}\right)}
\times \right. \nonumber \\ & &
\left. {{x_i (1-x_i^2)}\over{x_i^2 - \mu^2}} {{\partial
\sqrt{X}}\over{\partial x_i}} - {{2 m\mu}\over{x_i^2 - \mu^2}}{{d\ln X}\over{d\ln t}}
 \right\}.
\label{eq:di}
\ea
Here $\beta$ refers to the power-law index for $\rho_{\rm surf}$, or
that for $\rho$ near the surface. The benefit of the transformation
introduced in equation
\refnew{eq:psiform3} is that $d_i$ contains no derivatives on $\psi$.

Near the surface, $X = {\rm const}$, $d_i = 0$ as is expected and
equation \refnew{eq:psifull4} is separable. But in deeper region of
the planet, $X$ is a complex function of $x_1$ and $x_2$ and equation
\refnew{eq:psifull4} is not separable.

However, if $X$ is a smoothly varying function with a scale length
being the radius of the planet\footnote{In the case of Jupiter, $X =
1$ near the surface and gradually rises to larger values toward the
center.}, one can show that, in the WKB region, $d_i\psi_0 \sim
\psi_0$ and is $\sim 1/\lambda^2$ smaller than the ${\cal E}_i \psi_0$
term. So if we ignore the $d_i$ terms, we only introduce errors of
order ${\cal O}(\lambda^{-2}) \ll 1$. It is worth pointing out that
the non-WKB region occurs near the surface where our solution is
exact.

In conclusion, we can adopt the following approximate solution for
$\psi$
\be
\psi = \sqrt{X} \psi_0 = \sqrt{{\rho_{\rm surf}}\over{\rho}} 
\psi_1(x_1) \psi_2(x_2), 
\label{eq:psienv3}
\ee
with $\psi_i(x_i)$ satisfying ${\cal E}_i \psi_i + K^2 \psi_i = 0$.
This solution is exact near the surface and is accurate to ${\cal
O}(1/\lambda^2)$ in the WKB region. The latter attribute indicates
that the approximate solution describes accurately both the envelope
and the phase of the actual inertial-mode eigenfunction.

\section{Properties of Inertial-Modes}
\label{sec:property}

Having discussed methods to obtain inertial-mode eigenfunctions in
various density profiles, here we focus on studying general properties
of inertial-modes. This include the WKB properties, the normalization
relationship and density of modes in the frequency spectrum. Moreover,
we attempt to give readers a graphical impression of how
inertial-modes look like both inside and outside the planet. Lastly,
we compare inertial-modes against well known gravity- and
pressure-modes to gain intuition for this branch of eigenmodes.

\subsection{WKB Properties}
\label{subsec:wkb}

We first derive the WKB dispersion relation for inertial-modes. We
then determine the confine of the WKB region, and end with a general
derivation for the WKB envelope of inertial-mode amplitude.

In the WKB region, let $\bnabla \approx i \boldk $ and
$\partial/\partial z
\approx i k_z$, equation \refnew{eq:psifull} yields,
\be
k^2 - q^2 k_z^2 \approx 0,
\label{eq:kkz}
\ee
or $\mu \approx {{k_z}/{k}} \leq 1$. This is to be compared
with result from a more careful derivation (\S
\ref{subsubsec:dispersion}) which shows $\mu
\approx sin(n_2 \pi/\lambda)$. Since the $x_2$ axis
is largely along the $z$ axis (Fig. \ref{fig:x1x2}), $k_z \sim
n_2/R$. So we have $k \sim \lambda/R \sim 2 (n_1 + n_2)/R$, with $R$
being the planet radius (normalized to be one). Such a dispersion
relation implies that the mode frequency ($\omega = 2\Omega \mu$) does
not depend on the number of wiggles in a mode, but rather on the
direction of wave propagation. Modes that propagate close to the
rotation axis have higher frequencies ($\omega
\sim 2 \Omega$) than those that propagate close to the equator
($\omega \sim 0$).

Such a dispersion relation can be understood by the following physical
argument.  We relate $\boldxi$ to $\psi$ using equation
\refnew{eq:xipsi},
\be
\boldxi \approx i {{\boldk - q^2 {\boldez} k_z}\over{1-q^2}} \psi.
\label{eq:xipsi2}
\ee
So the fluid velocity ($\boldv = \partial \boldxi/\partial t$) is
perpendicular to the phase velocity ($\boldv_{\rm ph} = \omega/k^2
\boldk$), or $\boldv
\cdot \boldv_{\rm ph} \approx 0$. Inertial-waves are largely
transverse waves. A mode with its phase velocity along the equator
will show fluid motion in the $z$ direction. As a result, it
experiences little Coriolis force, and has a frequency that is close
to zero. In contrast, a mode with $\boldk $ along the $z$ direction
will experience the strongest restoring force and will have the
highest frequency.

While the phase velocity of the inertial-wave ${\boldv}_{\rm ph} =
\omega/k^2 \boldk$, the group velocity of the inertial-wave runs as
\ba
{\boldv}_{g} & = & \nabla_{\bold k} \omega = {{\partial
\omega}\over{\partial k_z}} \boldez +  {{\partial
\omega}\over{\partial k_h}} \boldeh = 2\Omega \left( {{k_h^2}\over{k^3}}\boldez - 
{{k_z k_h}\over{k^3}} \boldeh\right) \nonumber \\
& = & - {{\omega}\over {k^2}} (\boldk -
q^2 \boldez k_z),
\label{eq:vgroup}
\ea
where $\boldk = k_z \boldez + k_h \boldeh$. It is easy to show that
${\boldv}_{\rm ph} \cdot {\boldv}_g = 0$.

Now we search for the boundary that separates the WKB cavity from the
evanescent cavity for inertial-modes. Consider propagation in either
one of the $(x_1, x_2)$ directions. Equation \refnew{eq:krki2} shows
that the real part of the wave-vector remains fairly constant over the
whole planet, and is little affected by the density profile, while the
imaginary part of the wave-vector rises toward the surface as
$\beta/|x_i - \mu|$. The latter becomes comparable to the former when
$|x_i - \mu| \leq \beta/\lambda$. Recalling that $|x_i| = \mu$ occurs
at the surface, one sees that an upward propagating wave is reflected
near the surface at a depth $\delta \sim \beta/\lambda$ at most
latitudes, except near a special latitude $|\cos\theta| \sim \mu$
when the wave penetrates much higher into the envelope, to a depth of
$\delta \sim \beta^2/\lambda^2$. We call this special latitude where
$x_1 \sim |x_2| \sim \mu$ the ``singularity belt'' and it is an
important region for mode dissipation. We will return to this concept
in \S \ref{subsec:graphic}.


Lastly, we turn to study the WKB envelope of an
inertial-mode. Multiplying equation \refnew{eq:eqnmotion} by
$\dot{\boldxi}$, and simplifying the resulting expression using
equations
\refnew{eq:zerobrunt} \& \refnew{eq:ptorho}, we arrive at
the following equation of energy conservation,
\be
{\partial\over{\partial t}} \left( {\rho\over 2} \dot{\boldxi} \cdot
\dot{\boldxi} + {{p'^2}\over{2 \rho c_s^2}}\right) + 
\bnabla \cdot (p^\prime \dot{\boldxi}) = 0.
\label{eq:energyflux}
\ee
Terms in the first set of parenthesis are readily identified as the
energy density (including both kinetic energy density and
compressional energy density), while that in the second set is the
energy flux carried by the wave. Both quantities look identical to
those for non-rotating objects. This is expected since the Coriolis
force is an inertial force and does not do work or contribute to
energy.

The relative importance between the two energy density terms is,
\be
{{\omega^2 \rho |\boldxi|^2}\over{{p'}^2/\rho {c_s}^2}} = {{c_s^2
|\boldxi|^2}\over{\omega^2 \psi^2}} \approx {{c_s^2\lambda^2}\over{\omega^2
R^2}} \approx
\left({{c_s^2\lambda^2}\over{GM/R}}\right)\, 
\left({{GM/R^3}\over{\omega^2}}\right).
\label{eq:ratiodensity}
\ee
Since we are interested in $\lambda \gg 1$ modes, and since for
planets rotating well below the break-up speed, inertial-mode
frequency ($\omega \leq 2 \Omega$) is smaller than the fundamental
frequency of the planet ($\sqrt{GM/R^3}$), the above ratio is much
greater than unity over much of the planet. Exceptions occur near the
surface, where $\delta
\leq 1/\lambda^2 (\omega^2/GM/R^3) \ll 1/\lambda^2$, well
outside the WKB cavity. So the energy density of an inertial-mode is
dominated by its kinetic part. For a standing wave, the energy density
(and in this case, the kinetic energy density) remains constant in
time. So different velocity components must be out-of-phase with each
other. For instance, near the surface, $|\xi_\theta| \sim |\xi_\phi|
\gg |\xi_r|$, and $\xi_\theta$ and $\xi_\phi$ are $90\deg$ apart in
phase.

Take a standing wave of the form
\be
\boldxi (\boldr,t) = \boldxi(\boldr) e^{-i\omega t} + 
\boldxi^\star (\boldr) e^{i \omega t},
\label{eq:standing}
\ee
equation \refnew{eq:xipsi2} gives rise to the following energy density
(the time-independent part)
\ba
{\cal H} & = & {\rho\over 2} \dot{\boldxi} \cdot
\dot{\boldxi} + {{p'^2}\over{2 \rho c_s^2}}
 \approx 2{{\omega^2 }}\rho |\boldxi(\boldr)|^2 \nonumber\\
& \approx & 2 {{
\omega^2}} \, {{\boldk^2 - 2 q^2 k_z^2 + q^4 k_z^2}\over{(1-q^2)^2}}\,
\rho |\psi(\boldr)|^2
\nonumber \\
&\approx  & 2 {{\omega^2}}\, {{{\boldk}^2}\over{q^2-1}}\, \rho |\psi(\boldr)|^2 ,
\label{eq:Henergy}
\ea
while the energy flux 
\ba
\mbox{\boldmath${\cal F}$} & = & 
p^\prime \dot{\boldxi}= - 2 \omega\, {Im}[p^\prime(\boldr)
{\boldxi}^\star(\boldr)] = - 2 \omega^3 \rho\, {Im}[\psi\,
\boldxi^\star] \nonumber \\
& \approx & - 2 \omega^3 {{\boldk - q^2 {\boldez} k_z}\over{q^2-1}} 
 \rho |\psi(\boldr)|^2 = {\mbox{\boldmath $v_g$}} {\cal H},
\label{eq:energyf}
\ea
where ${\mbox{\boldmath $v_g$}}$ is the local group velocity as
derived in equation \refnew{eq:vgroup}. Inside the WKB propagating
cavity, 
the energy flux is constant (no reflection). This, coupled with the
fact that ${\boldk}$ is largely constant inside the planet, yields the
following WKB envelope
\be
|\psi(\boldr)| \propto {1\over{\sqrt{\rho}}}.
\label{eq:psienv4}
\ee
This agrees with results derived from more detailed considerations
(eqs. [\ref{eq:psienv1}], [\ref{eq:psienv2b}] \& [\ref{eq:psienv3}]).

\subsection{Mode Normalization}
\label{subsec:normalization}


We derive a scaling for the total energy in a mode.

\begin{figure*}
\centerline{\psfig{figure=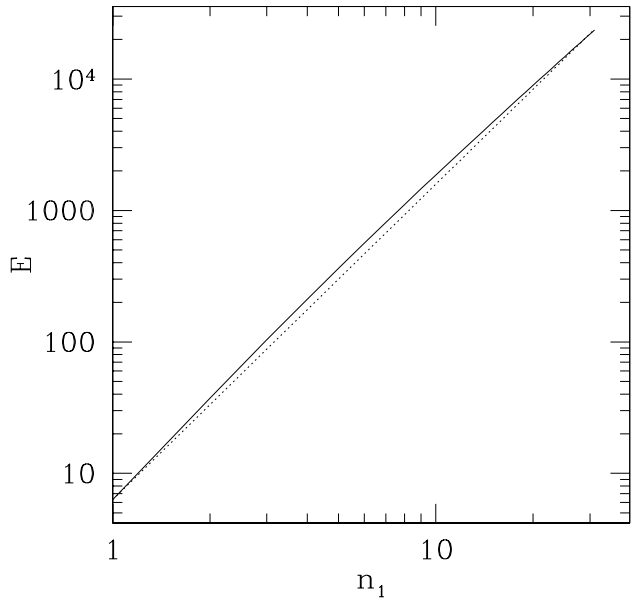,width=0.50\hsize}}
\caption{Mode energy as a function of $n_1$ for $m=-2$ even-parity 
inertial-modes in a $\beta=1$ model (solid curve). Here, all
inertial-modes are normalized with $\psi_2(x_2 = 0)=1$ and satisfy
$n_1 = n_2$ (so $\mu \sim 0.70$). The dotted curve is a fitting
power-law $E \propto n_1^{2.65}$.}
\label{fig:normalization}
\end{figure*}

Applying results from \S \ref{subsec:wkb} and the Jacobian defined in
equation \refnew{eq:jacobian}, we express total energy in a mode as
\ba
E & = & \int\, d^3 r\, {\cal H} \approx {\omega^2} \int\, d^3 r\, \rho
|\boldxi(\boldr)|^2  \nonumber \\
& \approx & - {{\omega^2}\over{1-q^2}}\, \int\, d^3 
r\, \boldk^2 \rho |\psi(\boldr)|^2 \nonumber \\
& \approx & {{\omega^2\mu}\over{(1-\mu^2)^2}} \, \int\,
(x_1^2 - x_2^2) dx_1 dx_2 d\phi\, \boldk^2 \rho |\psi|^2.
\label{eq:normal}
\ea
Most of the mode energy density lies inside the WKB region. Within
this region, each nodal patch ($\Delta x_1 \Delta x_2$) contributes a
comparable amount to the total energy. Since the total number of
patches is $\sim n_1 \times n_2$, we obtain
\ba
E & \sim & {{\omega^2\mu}\over{(1-\mu^2)^2}}\, n_1 n_2 \left.(\boldk^2
\rho |\psi|^2)\right|_{x_1\approx 1}^{x_2\approx 
0}\, \Delta x_1 \Delta x_2 \nonumber \\
& \sim & {{\omega^2\mu}\over{(1-\mu^2)^2}}\, n_1 n_2 
\left.(\rho |\psi|^2)\right|_{x_1\approx 1}^{x_2\approx 0}\,
\label{eq:normal2}
\ea
where the term $\rho |\psi|^2$ is supposed to be evaluated at the last
crest away from $x_1 = 1$ and $x_2 = 0$ (center of the
sphere). Adopting the arbitrary normalization of $\psi_2 (x_2 = 0) =
1$ and assuming that $\psi_1 (x_1 \approx 1)$ does not
depend on $n_1$, $E \propto n_1 n_2$ if evaluated at a fixed
frequency. Our numerical results, however, show
(Fig. \ref{fig:normalization}) that $E \propto n_1^{2.65}$ in a $\beta
= 1$ model when only $n_1 = n_2$ modes are considered (or $\mu \sim
0.70$).

The small difference results from the fact that as $n_1$ increases,
the amplitude of the last crest ($\psi_1 (x_1 \approx 1)$) increases
slightly. We adopt the numerical scaling $E \propto n_1^{2.65}$ for
our later studies.


\subsection{Density of Inertial-Modes}
\label{subsec:modedensity}

For our tidal problem, it is useful to study how dense inertial-modes
are within a given frequency interval. More specifically, we ask, for
a given frequency $\mu_0$, how close is the closest resonance
(smallest $|\mu-\mu_0|$) one can find among modes satisfying
wave-number $\lambda \leq \lambda_{\rm max}$.

\begin{figure*}
\centerline{\psfig{figure=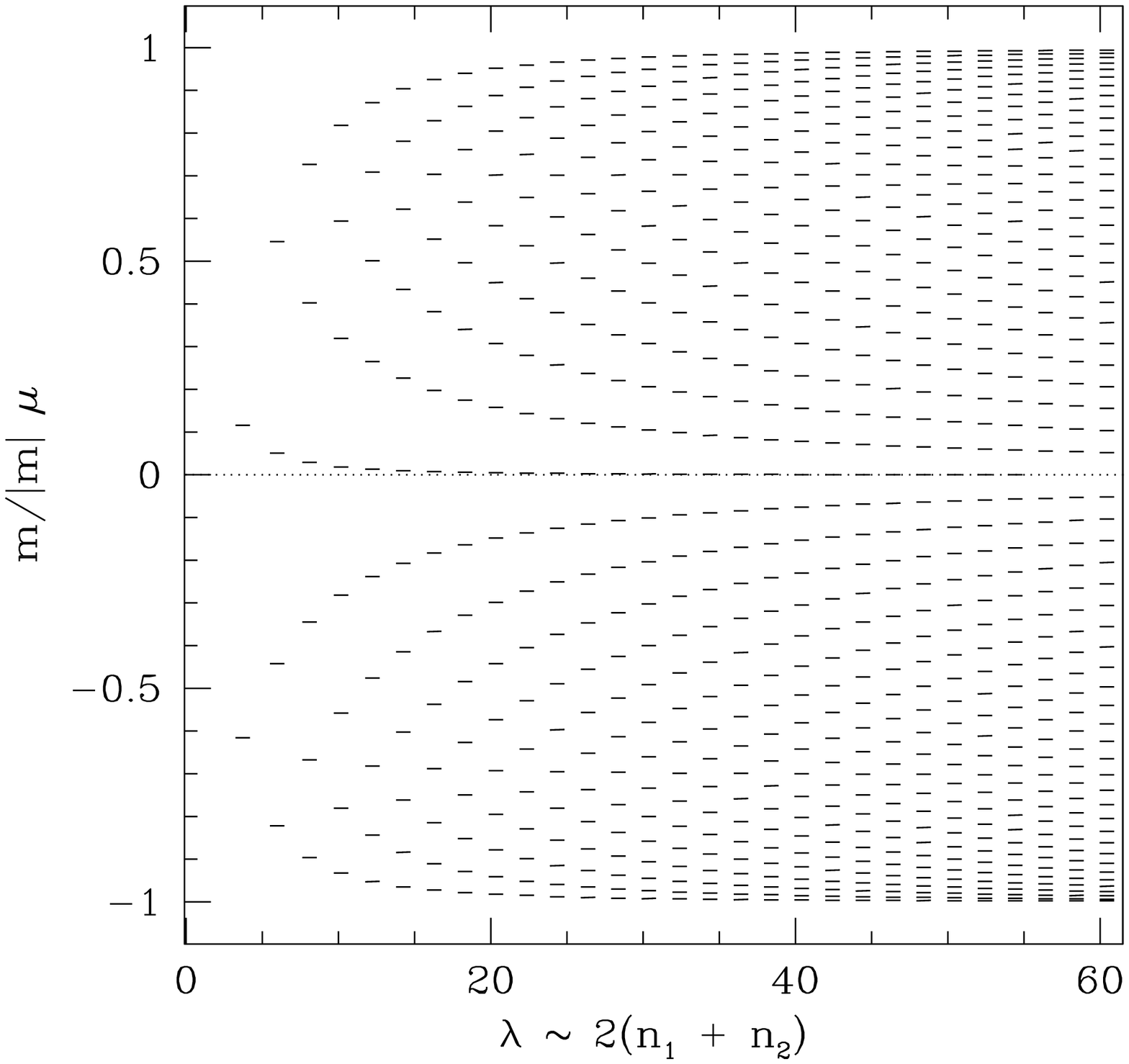,width=0.50\hsize}
}
\caption{Frequencies of even-parity, $|m| = 2$ inertial-modes 
(shown here as ${\mbox{SIGN}}\mu$) as a function of wave-number
$\lambda \sim 2 (n_1 + n_2)$. Modes above the dotted line are prograde
modes, and retrograde if below.  These frequencies are obtained using
a uniform density model but the overall features persist for more
general models. Note the asymmetry between the frequencies of prograde
and retrograde modes -- this follows from equations \refnew{eq:mueven}
\&
\refnew{eq:muodd}.}
\label{fig:mu-beta0-new}
\end{figure*}

Since frequencies of inertial-modes do not depend on the mode order,
but on the direction of mode propagation, modes of very different
wave-numbers can coexist in the same frequency
range. Fig. \ref{fig:mu-beta0-new} shows how $\mu$ depends on
$\lambda$ for even-parity, $|m|=2$ modes in a uniform density sphere.
The mean frequency spacing between modes of the same $\lambda$ (or
$n_1 + n_2$ for more general models) is $d\mu/dn_2 \approx
\sqrt{1-\mu^2}
\pi/\lambda$. This value decreases as $\mu$ approaches $1$ or as
$\lambda$ rises. The typical distance away from a resonance ($|\mu -
\mu_0|$) is half this value. When we include all modes with $\lambda < 
\lambda_{\rm max}$, the closest resonance likely has $|\mu - \mu_0| \approx
\sqrt{1-\mu^2} \pi/\lambda_{\rm max}^2 \sim \pi/\lambda_{\rm max}^2$.

\subsection{How do Inertial-Modes Look Like?}
\label{subsec:graphic}

In this section, we provide a graphical impression for how
inertial-modes look like both inside and on the surface of
planets. This may be helpful for readers as inertial-modes are
drastically different from modes in a non-rotating sphere, which can be
described by a product of a radial function and a single spherical
harmonic function.

\begin{figure*}
      \centering{
      \vbox{\psfig{figure=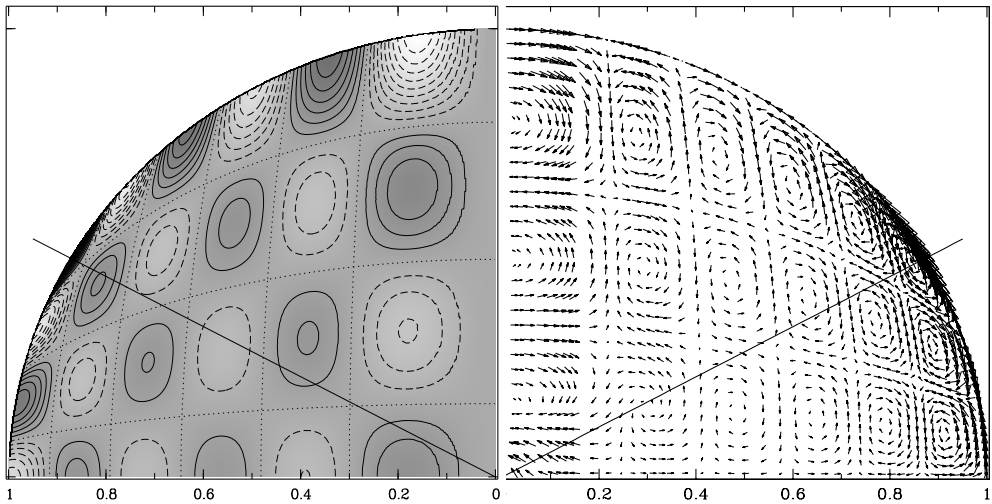,width=0.80\hsize,%
       bbllx=50pt,bblly=540pt,bburx=550pt,bbury=780pt,clip=}}
       }
\caption{A meridional look at the inertial-mode presented in Fig. 
\ref{fig:eigenfunction}. This mode is symmetric with respect to the equator,
retrograde ($m = -2$), with $n_1 = 5$, $n_2 = 3$. The planet model has
a density profile $\beta = 1$. The left panel shows the Eulerian
density perturbation ($\rho^\prime$), using both gray-scale and
equidistant contours. Lighter regions (and dashed contours) stand for
$\rho^\prime < 0$, while darker regions (solid contours) represent
$\rho^\prime > 0$. The dotted curves indicate the $(x_1, x_2)$
coordinates. Counting the number of nodes along each coordinate, one
recovers $n_1$ and $n_2$. The right panel shows the fluid velocity
($v_r$
\& $v_\theta$ components only) in the rotating frame as arrows, with
the size of the arrows proportional to $\sqrt{v}$.  Notice that $v_r$
vanishes at the surface, as is required by the boundary condition
(eq. [\ref{eq:dprho}]). Both the mode amplitude and the wave-vector
remain relatively constant over much of the planet, but rise sharply
toward the surface. This rise is most striking near the special angle
$|\cos\theta| = \mu = 0.4881$, marked here by straight lines.}
\label{fig:plota}
\end{figure*}

In the interior of a planet, an inertial-mode produces alternating
regions of compression and expansion, much like gravity- or
pressure-modes do, except that for inertial-modes, these regions are
lined-up along the $(x_1,x_2)$ coordinates. Fig. \ref{fig:plota}
depicts how the Eulerian density perturbation ($\rho^\prime$) and the
perturbation velocity in the rotating frame look like in a meridional
plane, for an example inertial-mode ($n_1 = 5, n_2 = 3$ and $m=
-2$). There are three noteworthy features. The first is that the
largest perturbations, as well as the steepest spatial gradients in
these quantities, are to be found near the surface, especially near
the angle $|\cos\theta| \approx \mu$. The second feature is that
velocity near the surface is purely horizontal. The radial component
vanishes as is required by the boundary condition (eq.
[\ref{eq:dprho}]). The third feature is that the velocity patterns
inside the planet take the form of vortex rolls. Inertial-modes
produce largely incompressible, and largely rotational motion
($|\nabla \times
\boldxi| \gg |\nabla \cdot \boldxi|$).

\begin{figure*}
\centerline{\psfig{angle=-90,figure=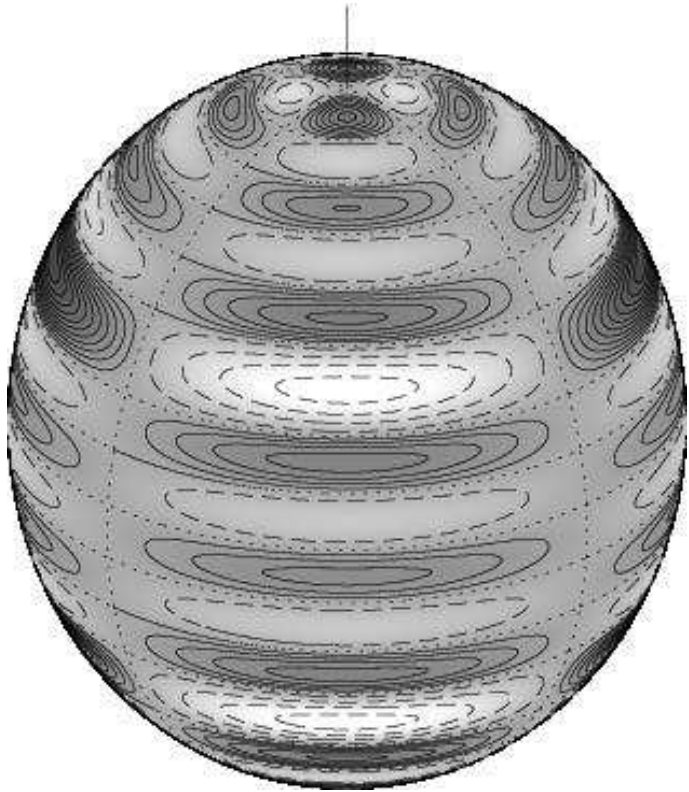,width=0.50\hsize}}
\caption{Surface density perturbation ($\rho^\prime$) for the same mode
($m = -2, n_1 = 5, n_2 = 3$) as in Fig. \ref{fig:plota}. The short
vertical stick marks the rotation axis. In both hemispheres, there is
a latitudinal belt within which density perturbation attains the
largest amplitude and varies with the steepest gradient. This lies
around $|\cos\theta| = \mu$ and we call it the ``singularity
belt''. The surface pattern of the mode also discloses the quantum
numbers: from the pole to the equator, there are $n_1$ nodal patches
above the singularity belt, and $n_2$ nodes below it.
}
\label{fig:density-surf}
\end{figure*}

Fig. \ref{fig:density-surf} shows a surface view of the density
perturbation for the same mode. For this retro-grade mode, the pattern
rotates retrogradely on the planet surface. First notice the number of
nodal patches on the surface. For a p- or g-mode in a non-rotating
star, the number of radial nodes does not show up in the surface
pattern. For inertial-modes, however, the values of $n_1$, $n_2$ and
$m$ (and even $\mu$) are all clearly embedded in the surface
pattern. The surface pattern tells all. Second notice the presence of
a belt (in both hemispheres) near $|\cos\theta| =
\mu$ where the mode exhibits both the largest
perturbation as well as the largest gradient of perturbation (see also
Fig. \ref{fig:plota}).  This we call the ``singularity belt'' and it
is a feature unique to inertial-modes. It is located at where both
$x_1 - \mu$ and $\mu-|x_2|$ $\leq 1/\lambda$, corresponding to a
region with a depth $\delta \sim 1/\lambda^2$ and an angular extent
$\delta\theta \sim 1/\lambda$.
This region will turn out to be very important for tidal
dissipation (Paper II).

\begin{figure*}
      \centering{
      \vbox{\psfig{figure=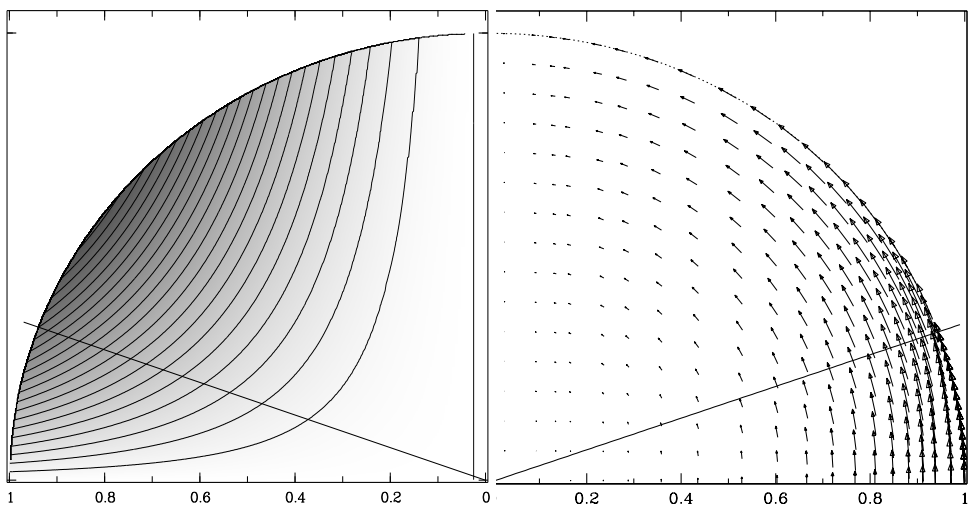,width=0.80\hsize,%
       bbllx=50pt,bblly=540pt,bburx=550pt,bbury=780pt,clip=}}
       }
\caption{As Fig. \ref{fig:plota}, except for
a $m=-2$ R-mode. This mode has $\mu = 1/(|m|+1) = 0.3333$ and in our
notation, $n_1 = n_2 = 0$ and $\ell = |m|+1 = 3$.  There is no
internal nodal point and the radial velocity is zero everywhere. This
mode exhibits odd symmetry toward the equator. }
\label{fig:plotb}
\end{figure*}

For comparison, we present a meridional look for the $m = -2$ R-mode
in Fig. \ref{fig:plotb}. R-modes are a special branch of
inertial-modes, and we discuss them in \S \ref{subsubsec:rmode}.

\subsection{Comparison with Gravity- and Pressure-modes}
\label{subsec:differing}

Modes restored by pressure or buoyancy (p- or g-modes) are familiar to
astronomers. To help understanding inertial-modes, we capitalize on
this familiarity by discussing the differences between these modes and
inertial-modes.

Inertial-modes are more analogous to g-modes than p-modes. Frequencies
of inertial-modes are higher if their direction of propagation is more
parallel to the rotation axis ($\omega/2 \Omega \sim k_z/k$, $z$ being
the rotation axis). Their frequencies are independent of the
magnitudes of the wave-vector and are constrained to $\omega < 2
\Omega$.  Similarly, higher frequency gravity-modes propagate more
parallel to the potential surface ($\omega/N \sim k_h/k$, $h$ being
the horizontal direction) satisfying $\omega < N$. Both inertial-waves
and g-waves are transverse waves, i.e., their group velocities
(direction of energy propagation) are perpendicular to their phase
velocities. P-modes, in comparison, experience stronger restoring
force and hence higher frequencies if their wavelengths are shorter.
Phase velocities and group velocities of p-modes are identical.

Due to their low-frequency nature, both g-modes and inertial-modes
cause little compression in their propagating cavities. As a result,
the energy of g-modes is dominated by the (gravitational) potential
and kinetic terms, which are alternately important for half a cycle,
each contributing in average half to the total energy. Inertial-modes
live in neutrally stratified medium, their energy is dominated by
kinetic energy alone. As a result, while g-modes (and p-modes) suffer
dissipation from radiative diffusion and viscosity, inertial-modes are
only sensitive to viscosity.

As p- or g-waves propagate toward the surface, their wavelengths
typically shorten, resulting in the largest dissipation. An
inertial-mode, however, propagates in its WKB cavity with a roughly
constant wavelength, except near the singularity belt ($r \sim R$ and
$\theta \sim \cos^{-1} \pm \mu$) where its wavelength shrinks
drastically. This is where we expect the largest dissipation to
occur.

Lastly, each p- or g-mode in a non-rotating star has an angular
dependence that is described by a single spherical harmonic function,
$P_\ell^m(\theta,\phi)$. In contrast, the angular dependence of each
inertial-mode is composed of a series of spherical harmonic
functions. This implies that while for p- and g-modes, only the $\ell
= 2$, $|m|=2$ branch can be excited by a potential force of the form
$P_2^2$ (such as the lowest order tidal force), for inertial-modes,
every $|m| = 2$ even-parity mode can be excited.  In this sense, the
frequency spectrum of inertial-modes is dense, and the probability of
finding a good frequency match (forcing frequency $\approx$ mode
frequency) is much improved over the non-rotating case (\S
\ref{subsec:modedensity}).

\section{Further Discussion}
\label{sec:discussion}

\subsection{Justifying Assumptions}
\label{subsec:assumption}

In our effort to obtain a semi-analytical solution for the
inertial-modes (\S \ref{sec:structure}), we have made a string of
simplifying assumptions. We justify them here.

%

In the analysis throughout this paper, we have ignored the
compressional term ($(1-q^2)\omega^2/c_s^2 \psi$ in eq.
[\ref{eq:psifull}]). This is also called the `anelastic approximation'
(ignoring $\rho^\prime$), and is often adopted when studying sub-sonic
flows in stratified medium. This assumption is justified by equation
\refnew{eq:ratiodensity} and discussions around it, where we show that 
the inertia term dominates over the compressional term in most of the
planet except well above the WKB propagating region.  Hence the
compressional term does not significantly affect either the frequency
or the structure of an inertial-mode.

Inertial-modes are excited by the tidal potential through its density
perturbation ($\rho^\prime$, see Paper II). Could we be removing tidal
forcing of inertial-modes by ignoring $\rho^\prime$? Fortunately, no.
Equation \refnew{eq:rhoprime} states that $\rho^\prime = \rho
\omega^2/c_s^2 \psi$. Since $\psi$ is non-zero, tidal forcing is  zero only 
when $c_s^2 \rightarrow \infty$. Taking the anelastic approximation
does not preclude tidal coupling. It does imply, however, that tidal
forcing of inertial-modes is expected to be weaker than tidal forcing
of the fundamental mode, by the same ratio that relates the inertia
term to the compressional term.

We have also ignored the potential perturbation caused by
inertial-modes themselves ($\Phi^\prime$ in
eq. [\ref{eq:eqnmotion}]). This so-called Cowling approximation is
justified here. The potential perturbation $\Phi^\prime$ is related to
the density perturbation by the Poisson equation,
\be
\nabla^2 \Phi^\prime = 4 \pi G \rho^\prime.
\label{eq:poisson}
\ee
Relative to the inertia term in equation \refnew{eq:eqnmotion}, the
potential term is smaller by
\be
\left|{{\rho \bnabla \Phi^\prime}\over{\rho\ddot{\boldxi}}}\right| \sim
{{4 \pi G \rho} \over {c_s^2 k^2}} \sim {1\over{\lambda^2}},
\label{eq:poisson2}
\ee
where we have used the results from equations \refnew{eq:rhoprime} and
\refnew{eq:xipsi2}. So the Cowling approximation is appropriate for 
high order modes ($\lambda \gg 1$), which are indeed the modes we are
concerned with (Paper II)

We have assumed that turbulent viscosity does not modify mode
structure significantly. This is equivalent of assuming that the
turbulent dissipation rate $\gamma \ll \omega$. This assumption is
confirmed by the numerical study to be presented in Paper II.

For numerical tractability, we have ignored the rotational deformation
of the planet, as well as the centrifugal force on the perturbed
motion. These introduce $\sim 10\%$ error in the case of Jupiter, and
much less for extra-solar planets. We do not expect the general
structure and dispersion relation of inertial-modes to be
significantly modified when these effects are taken into account.

Our last simplifying assumption concerns the planet's atmosphere. This
is by far our most uncertain assumption.  In solving for the
inertial-modes, we assume that the planet is fully convective (and
neutrally buoyant). However, Jupiter-like planets have surface
radiative zones with varying depths depending on their surface
composition and external irradiation. In the Jupiter model by
\citet{guillotreview}, the atmosphere is neutrally stratified up to
the
photosphere ($\sim 1$ bar in pressure), above which the temperature
profile is largely isothermal with a scale height ($\sim$ depth) $\sim
20 \km \sim 3\times 10^{-4} R$. Inertial-waves are reflected inward at
a depth $\delta \sim 1/\lambda$ at most latitudes, and at a much
shallower depth $\delta \sim 1/\lambda^2$ near the singularity belt
(\S \ref{subsec:wkb}). So modes with $\lambda \leq 1/{3\times
10^{-4}}\sim 3300$ will have most of their WKB cavity inside the
convective region and is only mildly affected by the isothermal layer;
while modes with $\lambda \leq 1/\sqrt{3\times 10^{-4}}\sim 60$ will
have their upper-most WKB turning point below this isothermal layer
and so will not be affected at all. In Paper II, we show that the
modes relevant for tidal dissipation have $\lambda \sim 60$.

Two concerns arise following the above discussion. 

The referee (Dave Stevenson) called our attention to the problem of
static stability in the Jovian atmosphere. This is currently uncertain
due to the presence of water condensation. A 'wet adiabat' is less
steep than a 'dry adiabat' as water in an upward moving parcel
condenses and releases latent heat. So the actual temperature profile
will be super-adiabatic for some (very moist) parcels and
sub-adiabatic for some (very dry) parcels.  
Such a 'conditional stability' makes it difficult to study the
propagation of inertial-modes.\footnote{Data gathered during the
descent of the Galileo probe \citep{seiff} showed that, at the entry
site, the temperature profile follows the dry adiabat from the
photosphere (1 bar) down to 16 bars, with some evidence (temperature
measurements and the presence of gravity waves) for a stable layer
from 15 to 24 bars.
%
It is now known that the probe entered a particularly dry area, so
this temperature profile may not be indicative of the whole
atmosphere.}

But if the stably stratified atmosphere does extend sufficiently deep,
and does harbor sufficiently strong buoyancy (\Bruntfreq $N
\geq \omega$), inertial-modes in the interior can couple to gravity-waves
in the atmosphere to form hybrid modes. Since most of the mode inertia
is in the interior, this will not affect much the frequencies of these
modes; however, it may affect their surface structure. Moreover,
inertial gravity-waves in the atmosphere may propagate upward freely,
break and dissipate at the low density environment in the upper
atmosphere. This brings about an extra damping mechanism for the
interior inertial-modes.

The second issue concerns extra-solar hot-jupiters. These planets are
strongly irradiated by their host stars. Their surface isothermal
layers may deepen to a depth of $\sim 10^{-2} R$. So even fairly
low-order inertial-modes may be affected.

For these two reasons, it is relevant to study behavior of
inertial-modes in the presence of an isothermal layer. We plan to do
so in the future.


\subsection{Special Case -- R-modes}
\label{subsubsec:rmode}

R-modes have been considered as a potential candidate for spinning
down young neutron stars and for emitting detectable gravitational waves
\citep[see, e.g.][]{owen}. So much attention have been paid to
this special class of inertial-modes (they also appear in Tables
\ref{tab:constantrho} \& \ref{tab:betarho}).
What is the relation between inertial-modes \citep[also called
'generalized R-modes' by][]{lindblom}) and R-modes?

R-modes are purely toroidal, odd-parity, retrograde
inertial-modes. They are, to the lowest order in $\Omega$,
incompressible and move only on spherical shells
\citep{papaloizou}. Setting $\xi_r = 0$ and $\divxi = 0$,
we find that the displacement vector for R-modes can be expressed
using a single stream function $Q = Q(r,\theta,\phi)$ where $\boldxi =
\bnabla \times (\boldr Q)$, or
\be
\xi_\theta = {{im}\over{\sin\theta}} Q, \hskip1.0in \xi_\phi = - 
{{\partial Q}\over{\partial \theta}}.
\label{eq:stream}
\ee
Taking the curl of equation \refnew{eq:xi1} and retaining only terms
to the lowest order in $\Omega$, we find
\be
{1\over{\sin\theta}} {\partial\over{\partial\theta}}\left(\sin\theta
{{\partial Q}\over{\partial \theta}}\right) - \left(mq +
{{m^2}\over{\sin^2\theta}}\right) Q = 0.
\label{eq:streameqn}
\ee
This is Legendre's equation with solution $Q \propto Y_{\ell^\prime,
m}(\theta,\phi)$ and eigenvalue $\mu =
\omega/2\Omega = - m/\ell^\prime(\ell^\prime+1)$ 
where $\ell^\prime \geq |m|$
\citep{papaloizou}. So the displacement vector $\boldxi = \bnabla
\times (\boldr Q) \propto \boldr \times \bnabla
Y_{\ell^\prime,m}$ and is purely axial (toroidal) in nature.
Obviously, both the angular dependence and the eigenfrequency of
R-modes are independent of the equation of state (compare R-mode
entries in Tables \ref{tab:constantrho} \& \ref{tab:betarho}). This is
expected as R-modes are restricted to move only along spherical shells.

According to this analysis, there should be infinite number of R-modes
for each $m$ value. However, in accordance with both \citet{lindblom}
and LF, we do not uncover any R-mode with $\ell^\prime > |m|$. This is
explained by LF \citep[see also][]{schenk}. They realize that under
the isentropic approximation,\footnote{ Isentropic refers to the fact
that both the background model and its adiabatic perturbation satisfy
the same equation of state. For instance, adiabatic perturbations in
neutrally buoyant models.}  poloidal-natured gravity-modes also have
frequencies $\sim {\cal O}(\Omega)$, besides from toroidal-natured
R-modes. So these two branches of modes are allowed to mix and this
eliminates the purely toroidal R-modes except for the lowest order
one, $\ell^\prime = |m|$. All other modes are a mixture of poloidal
(made of terms depending on $Y_{\ell^\prime,m}$ and $\nabla
Y_{\ell^\prime,m}$) and toroidal (made of terms depending on $\times
\bnabla Y_{\ell^\prime,m}$) terms. These are our general inertial-modes.
They can cause radial motion and their structure and their
eigenfrequencies depend on the equation of state.

Recall that we find eigenfunctions for uniform density planets to be
$\psi = \psi_1 (x_1) \psi_2 (x_2) \propto P_\ell^m(x_1)\,
P_\ell^m(x_2)$. When taking $\ell = |m| + 1$, we recover the above
R-mode solution.

Since R-modes are insensitive to the equation of state, they can exist
even in radiative stars \citep{papaloizou}. In contrast,
general inertial-modes exist only in neutrally buoyant medium.

\section{Summary}
\label{sec:summary}

In this work, we have studied inertial-modes with the purpose of
unraveling the role they may play in the tidal dissipation process of
Jupiter.

With the help of the ellipsoidal coordinates ($x_1, x_2$) first
adopted by \citet{bryan}, we have shown that the partial differential
equation governing inertial-modes in a sphere of neutrally buoyant
fluid can be separated into two ordinary differential equations when
the density is uniform or when the density has a power-law dependence.
The latter case is a novel result as far as we know.  For more general
density scalings, we show that we can obtain an approximate solution
to the inertial-modes
that is accurate to the second order in wave-vector. This important
result underlies much of our analytical study.

The dispersion relation $\mu = \omega/2\Omega \approx \sin(n_2
\pi/2(n_1+n_2)) $ (eq. [\ref{eq:muevenodd}]) rather generally 
describes how the frequency of an inertial-mode depends on its
structure. The quantum numbers $(n_1, n_2)$ are respectively the
number of nodes in the $x_1$ and $x_2$ coordinates, and the third
quantum number $m$ is the conventional azimuthal number. In our
notation, positive $m$ denotes prograde modes, and negative $m$
retrograde modes. So frequencies of inertial-modes depend on the
direction of wave propagation, with modes propagating close to the
rotation axis having higher frequencies. This dispersion relation also
indicates that inertial-modes are dense -- for any given frequency,
one can always find a combination of $n_1$ \& $n_2$ that approaches it
sufficiently closely.

We find that inertial-modes naturally cause small (but non-zero)
Eulerian density perturbations, of order $\omega \tau_{\rm
dyn}/\lambda$ (where $\tau_{\rm dyn}$ is the dynamical time-scale)
smaller than those by p-modes of comparable displacement amplitude
(eq. [\ref{eq:ratiodensity}]).
Their motion is nearly anelastic.
This implies that inertial-modes are only weakly coupled to the tidal
potential. It also implies that inertial-modes are not dissipated
primarily through heat diffusion, but rather through viscosity.

In its propagating region, an inertial-mode has a wave-vector $\boldk$
(measured in $x_1, x_2$ coordinates) that is nearly constant and is
insensitive to the local scale height and density distribution
(eq. [\ref{eq:krki2}]).  In this region, the amplitude envelope of an
inertial-mode rises with lowering of the density as
$1/\sqrt{\rho}$. An inertial-mode encounters its upper turning point
when the density scale-height becomes small and comparable to the
wave-vector. The depth of this turning point depends on the latitude:
when away from the spherical angle $\theta = \cos^{-1} \pm \mu$, the
turning point occurs at a depth $\sim R/\lambda$; while near this
angle, reflection occurs at a much shallower depth $\sim
R/\lambda^2$. Here, $\lambda \sim 2 (n_1 + n_2)$. We call the special
surface region around $|\cos\theta| = \mu$ the ``singularity
belt''. An inertial-mode has the highest amplitude as well as the
sharpest spatial gradient inside this belt. This region is associated
with the strongest turbulent dissipation.


Among the many simplifying assumptions we have adopted, the most
uncertain one concerns the static stability in the planet atmosphere
(\S \ref{subsec:assumption}). We plan to study its effects on
inertial-mode structure and dissipation in a future work.

\begin{acknowledgements}
Phil Arras has contributed a significant amount to the work presented
in this paper. This work would not have been possible without his
participation and I wish he has agreed to be a coauthor.
I thank him for a very enjoyable collaboration. I would also like to
acknowledge helpful discussions with J. Papaloizou, P. Goldreich,
G. Savonije, J. Goodman, Y. Levin and D. Lai over the years. Lastly,
this article benefited from the insightful comments by the referee,
Dave Stevenson.
\end{acknowledgements}

\begin{deluxetable}{ccrrrr}
\footnotesize
\tablecaption{Inertial-mode eigenfrequencies\tablenotemark{a}\,
 in a uniform density sphere.
\label{tab:constantrho}}
\tablewidth{0pt}
\tablehead{
\colhead{($\ell-|m|$)} & \colhead{parity\tablenotemark{b}} &
 \colhead{$|m|=1$}  & \colhead{$|m|=2$}   & \colhead{$|m|=3$}   & \colhead{$|m|=4$}
}
\startdata
 1\tablenotemark{c}
%
  &o &  -1.0000 &  -0.6667  &  -0.5000   & -0.4000  \\
 2&e &  -1.5099 & -1.2319  & -1.0532   & -0.9279   \\
  &e &  0.1766 & 0.2319  & 0.2532   & 0.2613 \\
 3&o &  -1.7080  & -1.4964  & -1.3402   & -1.2203   \\
  &o &  -0.6120  & -0.4669  & -0.3779   & -0.3175 \\
  &o &  0.8200 & 0.7633  & 0.7181   & 0.6807 \\
 4&e &  -1.8060 & -1.6434  & -1.5119   & -1.4044   \\
  &e &  -1.0456  & -0.8842  & -0.7735   & -0.6920    \\
  &e &  0.0682  & 0.1018  &  0.1204  & 0.1312  \\
  &e &  1.1834 & 1.0926  & 1.0222   & 0.9652 \\
 5&o &  -1.8617  & -1.7340  & -1.6236   & -1.5290   \\
  &o &  -1.3061  & -1.1530  & -1.0401   & -0.9525   \\
  &o &  -0.4404  & -0.3595  & -0.3040   & -0.2635 \\
  &o &  0.5373 & 0.5100  & 0.4869   & 0.4669 \\
  &o &  1.4042 & 1.3080     & 1.2309   & 1.1670 \\
\enddata
\tablenotetext{a}{Here, we present eigenfrequencies 
as $2\mu m/|m| = {\mbox{SIGN}}[m]\,\omega/\Omega$, where $\omega$ is
the eigenfrequency in the rotating frame, and $\Omega$ the rotational
frequency. So positive values denote prograde modes, while negative
ones retrograde (opposite to that in LF). The eigenvalues $\ell$ and
$m$ appear in the eigenfunction as $\psi = P_\ell^m(x_1)
P_\ell^m(x_2)$. For each $(\ell,|m|)$ pair, there are $(\ell-|m|)$
distinct eigenfrequencies. These results agree with those obtained by
LF.}
\tablenotetext{b}{This denotes the parity with respect to the 
equatorial plane: e for even and o for odd.
}
\tablenotetext{c}{This row shows the eigenfrequencies of 
pure r-modes. They are odd-parity, retrograde modes satisfying
$\omega=2\Omega/(|m|+1)$ to the first order in $\Omega$.}
\end{deluxetable}

\begin{deluxetable}{ccrrrrrrr}
\footnotesize
\tablecaption{Eigenfrequencies\tablenotemark{a}\,
 of $|m| = 2$ inertial-modes for various power-law density profiles
\label{tab:betarho}}
\tablewidth{0pt}
\tablehead{
\colhead{($\ell-|m|$)} & \colhead{parity} &
\colhead{$\beta=0.0$}   & \colhead{$\beta=0.1$}  & \colhead{$\beta=0.5$} 
  & \colhead{$\beta=1.0$} &  \colhead{$\beta=2.0$} 
& \colhead{$p= k \rho^2$(LF)\tablenotemark{b}}}
\startdata
 1\tablenotemark{c}
  &o & -0.6667	& -0.6667 & -0.6667 & -0.6667 & -0.6667 & -0.6667\\
 2&e & -1.2319	& -1.2133 & -1.1607 & -1.1224  & -0.8628 & -1.1000 \\	
  &e & 0.2319	& 0.2673 & 0.3830 & 0.4860  & 0.6159 	& 0.5566\\
 3&o & -1.4964	&-1.4789 &-1.4256&-1.3822&-1.3317	& -1.3578	\\
  &o & -0.4669  &-0.4732&-0.4924&-0.5082&-0.5270	& -0.5173	\\
  &o &0.7633	& 0.7942&0.8919	&0.9761	&1.0798		& 1.0259	\\
 4&e &-1.6434	&-1.6287&-1.5820&-1.5415&-1.4909	& -1.5196	\\
  &e &-0.8842	&-0.8811&-0.8726&-0.8671&-0.8628	& -0.8629	\\
  &e &0.1018	&0.1188	&0.1780	&0.2364	&0.3199		& 0.2753	\\
  &e &1.0926	&1.1137	&1.1814	&1.2408	&1.3150		& 1.2729	\\
 5&o &-1.7340	&-1.7217&-1.6820&-1.6460&-1.5988	& -1.6272	\\
  &o &-1.1530	&-1.1471&-1.1286&-1.1133&-1.0954	& -1.1044	\\
  &o &-0.3595	&-0.3669&-0.3904&-0.4108&-0.4367	& -0.4217	\\
  &o &0.5100	&0.5313	&0.6020	&0.6673	&0.7554		& 0.7039	\\
  &o &1.3080	&1.3226	&1.3700	&1.4122 &1.4654		& 1.4339	
\enddata
\tablenotetext{a}{Eigenfrequencies are again presented 
as $2\mu\, m/|m| = {\mbox{SIGN}}[m]\, \omega/\Omega$. with positive
values denoting prograde modes, and negative ones retrograde modes.
Here we define $\ell = 2(n_1 + n_2) + |m| - \delta$, where $\delta =
0$ for
even-parity and $1$ for odd-parity, and $n_1$, $n_2$ are the number of
nodes in $x_1$ and $x_2$ ranges, respectively. When $\beta = 0$, this
$\ell$ matches that appearing in the Legendre polynomial ($P_\ell^m$).
The number of eigenmodes is conserved when $\beta$ varies, with a
close one-to-one correspondence between modes in different density profiles.}
\tablenotetext{b}{For comparison, we list results calculated by
LF (their Table 5) using a series expansion method for a polytrope
model $p \propto \rho^2$. Not surprisingly, their eigenfrequencies
fall somewhere in between our results for $\beta=1$ and $\beta=2$
models.}
\tablenotetext{c}{This row shows pure r-modes. Both the frequency 
and the eigenfunction of these modes do not depend on the equation of
state, to the lowest order in $\Omega$.}
\end{deluxetable}

\begin{figure*}
\centerline{\psfig{figure=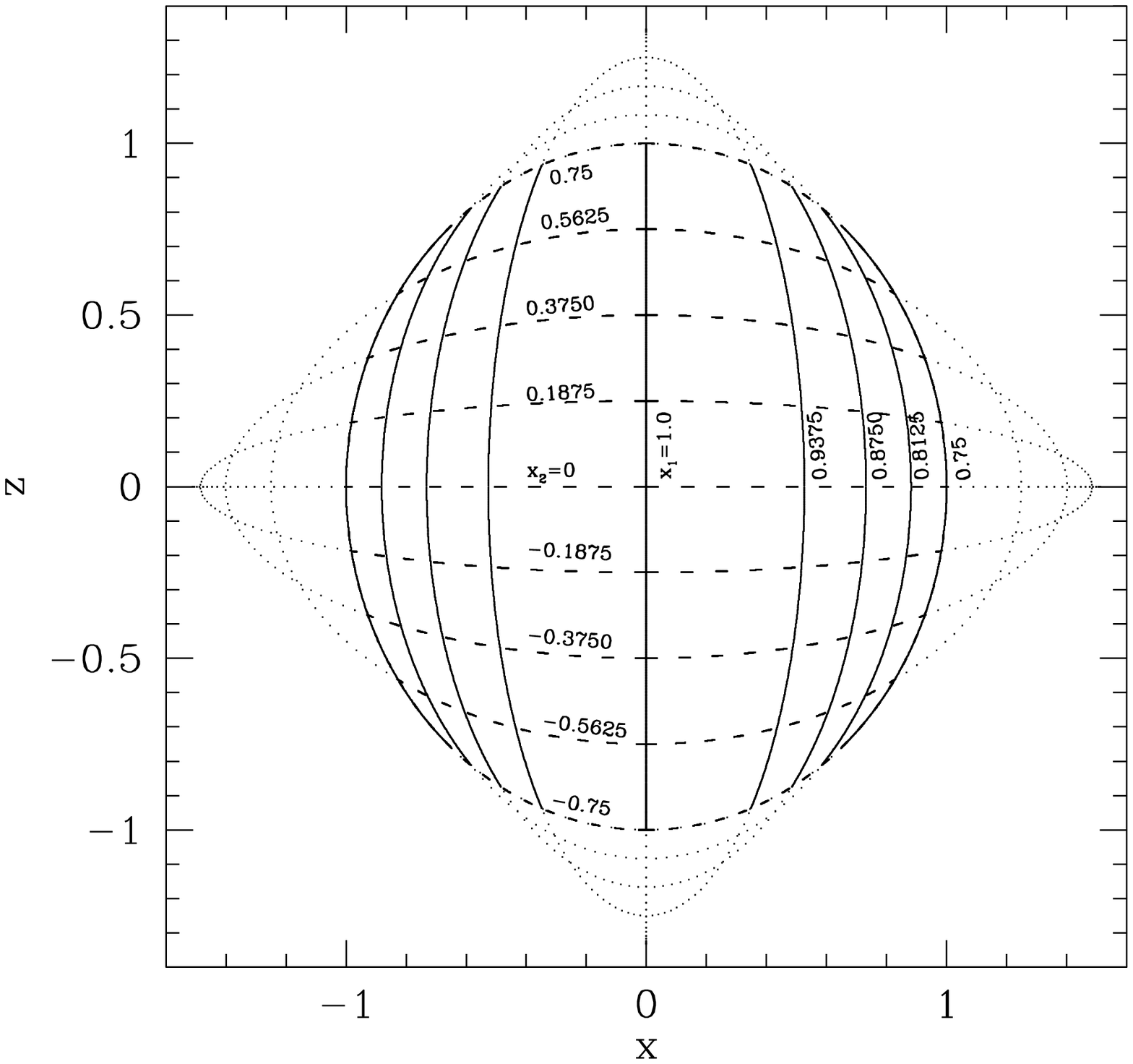,width=0.50\hsize}
\psfig{figure=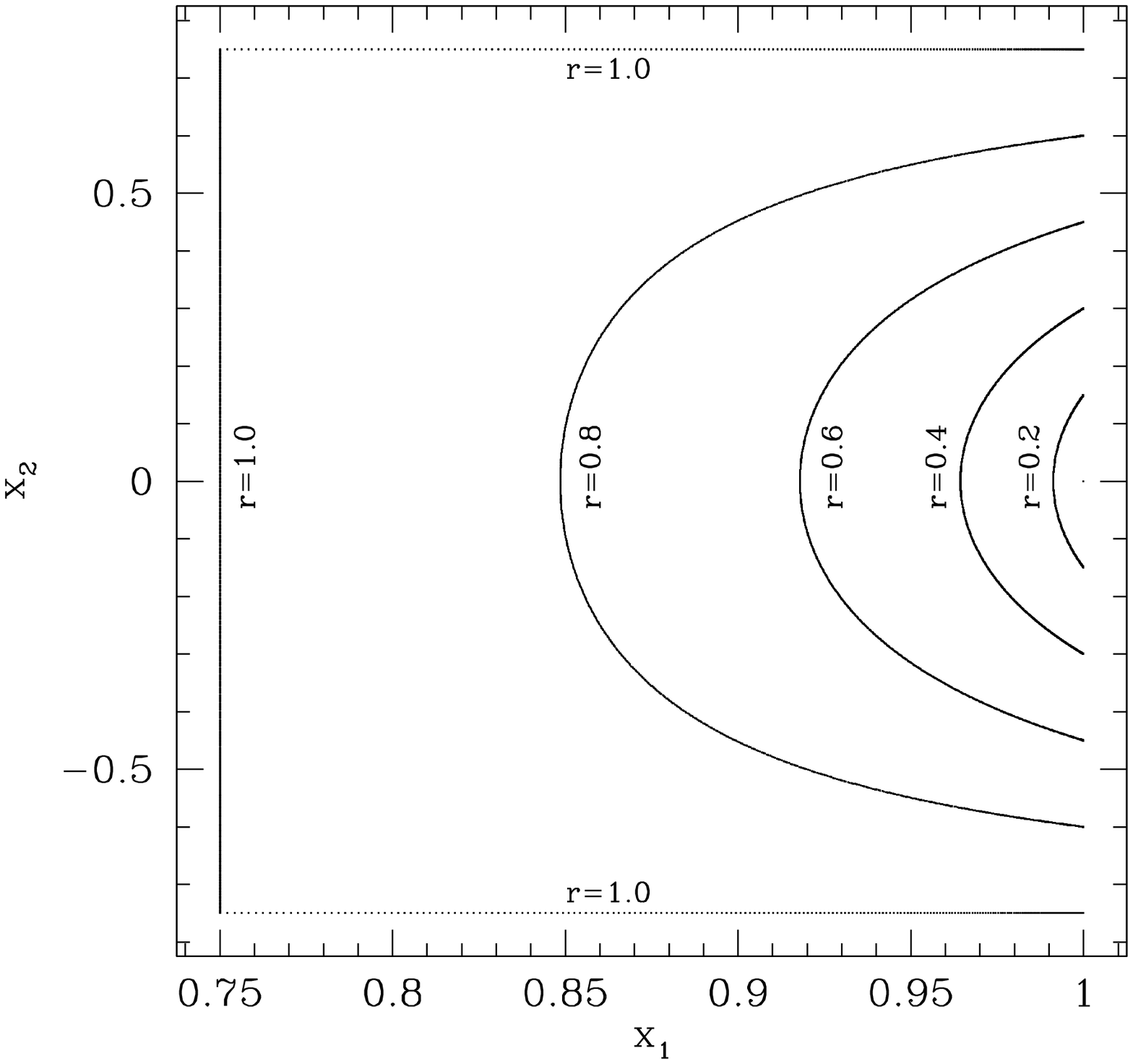,width=0.50\hsize}
}
\caption{
The ellipsoidal coordinates for $\mu = 0.75$.  Left panel:
equidistant curves of constant $x_1$ (solid curves) and constant $x_2$
(dashed curves) in a meridional plane.
They become dotted curves outside the surface of the planet, which
corresponds to either $x_1 = \mu$ or $|x_2| = \mu$ or both (at the
special latitude $|\cos \theta| = \mu$).  The equidistant curves are
most closely packed near the surface at this special latitude.  Over
the rest of the sphere, the $x_1$ axis runs similarly as the
cylindrical radius, while the $x_2$ axis runs largely parallel to the
rotation axis.  Right panel: equidistant curves of constant spherical
radius in the $x_1 - x_2$ plane.  
}
\label{fig:x1x2}
\end{figure*}

\begin{appendix}

\section{Ellipsoidal Coordinates}
\label{sec:coordinate}

\citet{bryan} introduced a new set of coordinates under which
the left-hand-side of equation \refnew{eq:psifull} becomes
separable. For any given density profile, the right-hand-side of the
same equation is generally inseparable -- however, for the two special
cases of uniform and power-law density profiles (\S
\ref{subsec:constantrho} and \ref{subsec:powerlaw}), the 
right-hand-side becomes separable and we can obtain (semi)-analytical
solutions for the inertial-modes. This new set of coordinates allow us
to obtain good approximate solutions for more general density profiles
(\S \ref{subsec:approximate}).

We scale all length in the problem by the radius of the planet so that
the Cartesian coordinates, $x$, $y$ and $z$, fall within the range of
$(-1,+1)$. The new ellipsoidal coordinates $(x_1, x_2, \phi)$ are
related to the Cartesian coordinates as
\begin{eqnarray}
x & = & \left[ {{(1-x_1^2)(1-x_2^2)}\over{1-\mu^2}}\right]^{1\over 2} \cos\phi,
\nonumber \\
y & = & \left[ {{(1-x_1^2)(1-x_2^2)}\over{1-\mu^2}}\right]^{1\over 2} \sin\phi,
\nonumber \\
z & = & {{x_1 x_2}\over{\mu}}, \label{eq:x1x2appear}
\end{eqnarray}
with $x_1 \in [\mu,1]$, $x_2 \in [-\mu,\mu]$, and $\phi$ is the usual
azimuthal angle with $\phi \in [0,2\pi]$. The cylindrical and
spherical radii are given by
\begin{eqnarray}
\pomega^2 & = & x^2 + y^2 = {{(1-x_1^2)(1-x_2^2)}\over{({1-\mu^2})}},\nonumber \\
r^2 & = & x^2 + y^2 + z^2 = 1 - {{(x_1^2 -
\mu^2)(\mu^2-x_2^2)}\over{(1-\mu^2)\mu^2}}.
\label{eq:pomegar}
\end{eqnarray}

The Jacobian that relates the new coordinate system $(x_1, x_2, \phi)$
to the Cartesian $(x,y,z)$ is
\begin{eqnarray}
\begin{array}{c}
{\cal J} = {\rm det} \left| \begin{array}{lll}
{{\partial x}/{\partial x_1}}\hskip0.1in & 
{{\partial x}/{\partial x_2}} \hskip0.1in & 
{{\partial x}/{\partial \phi}}\\
{{\partial y}/{\partial x_1}}\hskip0.1in& 
{{\partial y}/{\partial x_2}} \hskip0.1in& 
{{\partial y}/{\partial \phi}} \\
{{\partial z}/{\partial x_1}} \hskip0.1in& 
{{\partial z}/{\partial x_2}} \hskip0.1in & 
{{\partial z}/{\partial \phi}} \\
\end{array}\right|  = {{x_1^2 - x_2^2}\over{(1-\mu^2)\mu}} 
\end{array}
\label{eq:jacobian}
\end{eqnarray}
so the volume element $dx\, dy\, dz = {\cal J} dx_1\, dx_2\, d\phi$.

Fig. \ref{fig:x1x2} depicts the equi-distance curves of $(x_1, x_2)$
in a meridional plane, as well as the equi-distance curves of radius
$r$ in the $x_1 - x_2$ plane. In 3-D, surfaces of constant $x_1$
appear as co-axial cylinders around the $z$-axis (or prolate
ellipsoids), while surfaces of constant $x_2$ resemble bandannas
symmetrical with respect to the equator (or oblate ellipsoids). 
On the spherical surface, either $x_1$ or $|x_2|$ (or both) equals
$\mu$ -- we call the region where $x_1 \approx |x_2| \approx
\mu$ the 'singularity belt', a region of special significance. 
Moreover, $x_1 = 1$ at the rotation axis and $x_2 = 0$ at the equator.

Partial differentiation with respect to $\pomega$, $z$ and $r$ can be
expressed in the new coordinates as,
\begin{eqnarray}
\left.{{\partial}\over{\partial \pomega}}\right|_z & 
= & \left.{{\partial x_1}\over{\partial
\pomega}}\right|_z \left.{{\partial}\over{\partial x_1}}\right|_{x_2} + 
\left.{{\partial x_2}\over{\partial \pomega}}\right|_z 
\left.{{\partial }\over{\partial x_2}}\right|_{x_1} 
 =  - {{(1-\mu^2)\pomega}\over{x_1^2 -
x_2^2}}\left(x_1 {{\partial}\over{\partial x_1}} - x_2
{{\partial}\over{\partial x_2}}\right) \label{eq:partialpomega}\\
\left.{{\partial}\over{\partial z}}\right|_{\pomega} 
& = & \left.{{\partial x_1}\over{\partial z}}\right|_{\pomega}
\left.{{\partial}\over{\partial x_1}}\right|_{x_2} + \left.{{\partial
x_2}\over{\partial z}}\right|_{\pomega} \left.{{\partial
}\over{\partial x_2}}\right|_{x_1} = 
{{\mu}\over{x_1^2 - x_2^2}}\left[x_2 (x_1^2-1)
{{\partial}\over{\partial x_1}} - x_1 (x_2^2-1)
{{\partial}\over{\partial x_2}}\right],
\label{eq:partialz}\\
\left.{{\partial}\over{\partial r}}\right|_{\theta} 
& = & \left.{{\partial r}\over{\partial z}}\right|_{\theta}
\left.{{\partial}\over{\partial z}}\right|_{\pomega} + \left.{{\partial
r}\over{\partial \pomega}}\right|_{\theta} \left.{{\partial
}\over{\partial \pomega}}\right|_{z} =
- {{(1-x_1^2)x_1}\over{(x_1^2- x_2^2) r}} {\partial\over{\partial x_1}}
+ {{(1-x_2^2)x_2}\over{(x_1^2- x_2^2) r}} {\partial\over{\partial x_2}}.
\label{eq:partialr}
\end{eqnarray}

\section{Making $\beta$ Models}
\label{sec:betamodel}

We construct a hydrostatic model for a power-law density profile of
the form $\rho = (1-r^2)^\beta$ (eq. [\ref{eq:powerrho}]). Any value
of $\beta$ is allowed, in contrast to the polytrope case where the
polytrope index has to be smaller than $5$.

For our numerical integration, we adopt $\ln \rho $ as the independent
variable, and $\ln p$ and $M(r)$ as the variables. Here $p$ is
pressure and $M(r)$ is the mass within radius $r$. The normal
hydrostatic equations apply with the boundary conditions that at the
center $M(r) = 0$ and at the surface,
\be
p = {{2^\beta g_{\rm surface}}\over{\beta+1}} (1-r)^{\beta+1},
\label{eq:psurface}.
\ee
where surface gravity $g_{\rm surface} = GM(r=1)/r(r=1)^2$.  This is
obtained using $dp/dr = - \rho g$.  We can then solve the boundary
value problem to obtain the interior structure of such a model. Here,
$r$ runs from $0$ to $1$ and the density scale is arbitrary.

\end{appendix} 

\end{document}